\documentclass[journal,twocolumn,font=12]{IEEEtran}
\ifCLASSINFOpdf
   \usepackage[pdftex]{graphicx}
\else
   \usepackage[dvips]{graphicx}
\fi
\usepackage[cmex10]{amsmath}
\usepackage{amssymb}
\usepackage{xcolor}
\usepackage{bbm}

\allowdisplaybreaks

\usepackage{array}

\usepackage{cite}

\newtheorem{theorem}{Theorem}
\newtheorem{example}{Example}
\newtheorem{definition}{Definition}

\newtheorem{proposition}{Proposition}
\newtheorem{corollary}{Corollary}
\newtheorem{remark}{Remark}
\newcommand{\tp}{\textnormal{tp}}

\renewcommand{\and}{\textnormal{and}}

\newcommand{\mw}[1]{{\color{black}#1}}

\newcommand{\nobinning}{\textnormal{nobin}}
\newcommand{\binning}{\textnormal{bin}}
\newcommand{\simple}{\textnormal{dcpled}}
\newcommand{\binsimple}{\textnormal{bin,dcpled}}

\newcommand{\YC}{Y_{\textnormal{C}}}
\newcommand{\YH}{Y_{\textnormal{H}}}
\newcommand{\ZC}{Z_{\textnormal{C}}}
\newcommand{\ZH}{Z_{\textnormal{H}}}
\newcommand{\ZCt}{Z_{\textnormal{C},t}}
\newcommand{\YCt}{Y_{\textnormal{C},t}}
\newcommand{\ZHt}{Z_{\textnormal{H},t}}
\newcommand{\YHt}{Y_{\textnormal{H},t}}
\newcommand{\ZCrest}{Z_{\textnormal{C},t+1}^n}
\newcommand{\YCrest}{Y_{\textnormal{C},t+1}^n}

\begin{document}

%
\title{Hypothesis Testing over the Two-Hop Relay Network}


\author{Sadaf Salehkalaibar, \emph{IEEE Member}, Mich\`ele Wigger, \emph{IEEE Senior Member}, Ligong Wang, \emph{IEEE Member}
	\thanks{S.~Salehkalaibar is  with the Department of Electrical and Computer Engineering, College of Engineering, University of Tehran, Tehran, Iran, s.saleh@ut.ac.ir,}
	\thanks{M.~Wigger is with   LTCI,  Telecom ParisTech, 75013 Paris, michele.wigger@telecom-paristech.fr,}
	\thanks{L. Wang is  with  ETIS, Universit\'e Paris Seine, Universit\'e de Cergy-Pontoise, ENSEA, CNRS, ligong.wang@ensea.fr.}
	\thanks{M. Wigger was supported by the European Research Council
under Grant 715111. This paper was presented in part at the 2017 IEEE
Information Theory Workshop.}
	
}
\maketitle

\begin{abstract}

Coding and testing schemes  and the corresponding achievable type-II error exponents  are presented for binary hypothesis testing over two-hop relay networks. The schemes are based on cascade source coding techniques and {unanimous decision-forwarding}, the latter meaning that a terminal decides on the null hypothesis only if all previous terminals have decided on the null hypothesis. If the observations at the transmitter, the relay, and the receiver form a Markov chain in this order, then, without loss in performance, the proposed cascade source code can be replaced by two independent point-to-point source codes, one for each hop. The decoupled scheme (combined with decision-forwarding) is shown to attain the optimal type-II error exponents for various instances of ``testing against conditional independence.'' The same decoupling is shown to be optimal also for some instances of ``testing against independence,'' when the observations at the transmitter, the receiver, and the relay form a Markov chain in this order, and when the relay-to-receiver link is of sufficiently high rate. 
For completeness, the paper also presents an analysis of the Shimokawa-Han-Amari binning scheme for the point-to-point hypothesis testing setup.	
\end{abstract}



%
\IEEEpeerreviewmaketitle

\section{Introduction}
As part of the Internet of Things (IoT), sensor applications are rapidly increasing, thanks to lower cost and better performance of sensors.
One of the major  theoretical challenges in this respect is sensor networks with multiple sensors  collecting correlated data, which they communicate to one or multiple decision centers. Of special practical and theoretical interest  is to study the tradeoff between the quality of the decisions taken at the centers and  the required  communication resources. In this work, following the approach in  \cite{Csiszar86, Han}, we consider problems where decision centers have to decide on a binary hypothesis $\mathcal{H}=0$ or $\mathcal{H}=1$ that determines the underlying joint probability mass function (pmf) of all the terminals' observations. Our goal is to characterize the set of possible \emph{type-II error exponents} (i.e., the error exponent in deciding $\hat{\mathcal{H}}=0$ when in fact $\mathcal{H}=1$) as a function of the available communication rates such that the \emph{type-I error probabilities} (i.e., error probabilities of deciding  $\hat{\mathcal{H}}=1$ when in fact $\mathcal{H}=0$) vanish as the lengths of the observations grow.
Previous works on this \emph{exponent-rate region}  considered  communication scenarios over dedicated noise-free links from one or many transmitters to a single decision center \cite{Csiszar86,Amari,Wagner}  or from a single transmitter to two decision centers \cite{Michele, Roy,PierreISIT}. The hypothesis testing problem from a signal processing perspective has been studied in several works \cite{Varshney, Veer, Tara, Poor,Nanda}. Recently, simple interactive communication scenarios were also considered \cite{Lai1,Lai2,Kim}, as well as hypothesis testing over \emph{noisy} communication channels \cite{Gunduz,Michele,Sadaf-IT}.  All these distributed hypothesis testing problems are open in the general case; exact solutions have only been found for instances of ``testing against independence'' \cite{Csiszar86} and of ``testing against conditional independence'' \cite{Wagner}. ``Testing against independence'' refers to a scenario where the observations' joint pmf under $\mathcal{H}=1$ is the product of the marginal pmfs under $\mathcal{H}=0$, and ``testing against conditional independence'' refers to a scenario where this independence holds only conditional on some sub-sequence that is observed at the receiver and that has the same joint distribution with the sensor's observations under both hypotheses. 

\begin{figure}[t]
		\centering
		\includegraphics[scale=0.27]{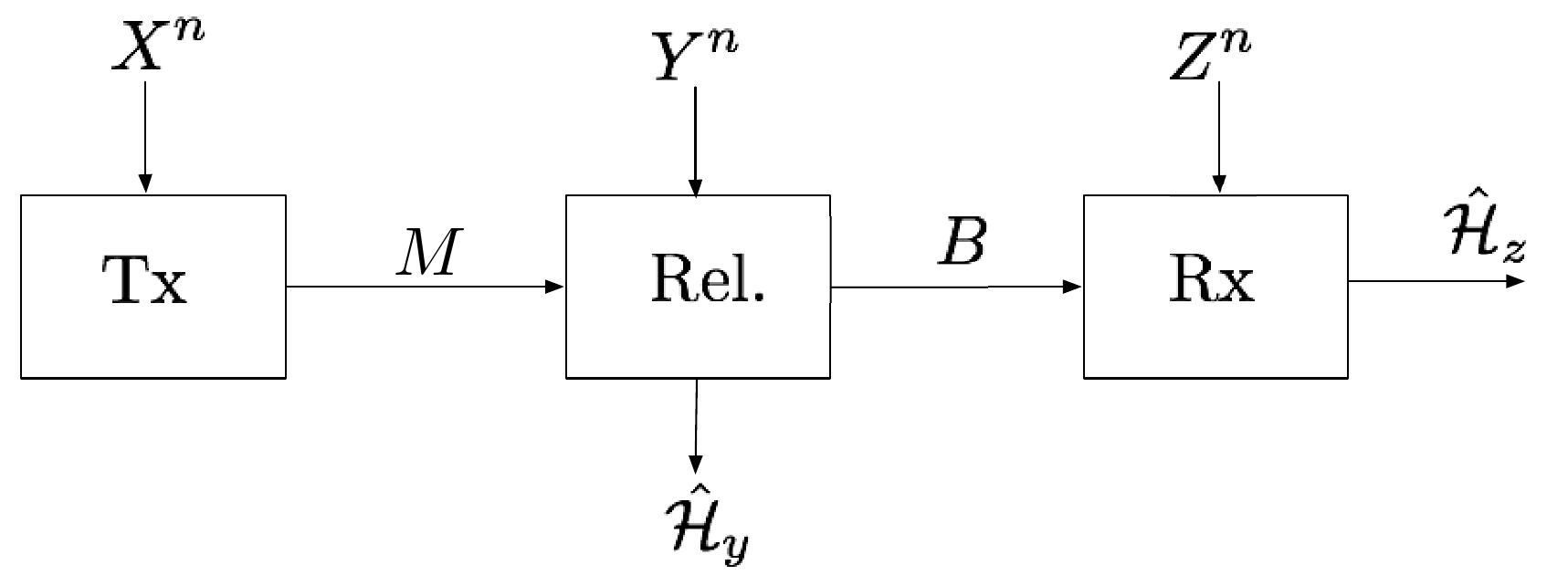}
		\caption{Hypothesis testing over a single-relay multi-hop channel}
		\label{cascade}
	\end{figure}

The focus of this paper is on the \emph{two-hop network} depicted in Fig.~\ref{cascade}. We model a situation with three sensors and two decision centers. The first terminal (the transmitter) models a simple sensor that observes an $n$-length sequence $X^n$. The second terminal (the relay) includes both a sensor observing the $n$-length sequence $Y^n$ and a decision center which produces the guess $\hat{\mathcal{H}}_y\in\{0,1\}$. Similarly, the third terminal (the receiver) includes a sensor observing $Z^n$ and a decision center producing the guess  $\hat{\mathcal{H}}_z\in\{0,1\}$. Communication is directed and in two stages. The transmitter communicates directly with the relay over a noise-free link of rate $R>0$, but it cannot directly communicate with the receiver, e.g.,  because the receiver is too far away. Such a restriction is particularly relevant for modern IoT applications where sensors are desired to consume very little energy so as to last for decades without the batteries being replaced.  On the other hand, the receiver is assumed  to be sufficiently close to the relay so that the relay  can communicate directly with it over a  noise-free link of rate $T>0$. The task of the relay is not only to communicate information about its own observation to the receiver but also to \emph{process and forward} information that it receives from the transmitter. Two-hop networks have previously been studied in information theory for source coding or coordination. These problems are open in general. Solutions to special cases were presented in \cite{Yamamoto,CuffSu, Diggavi,Cuff-coordination, Tandon, Chia,Weissman,Ahmadi}.

	In this paper, we propose two coding and testing schemes for binary hypothesis testing over the two-hop relay network. The two schemes apply two different 
		 source coding schemes  for the two-hop relay network to convey quantization information about the distributed observations to the relay and the receiver, and combine these schemes with 
		  a \emph{unanimous decision-forwarding} strategy. In this latter strategy, each of the terminals tests whether its reconstructed source sequences are jointly typical with its own observation under the null hypothesis $\mathcal{H}=0$. If the test is positive and the 
	preceding terminals have also decided on $\hat{\mathcal{H}}=0$, then the terminal declares this null hypothesis $\hat{\mathcal{H}}=0$. Otherwise it declares the alternative hypothesis $\hat{\mathcal{H}}=1$. In both cases, it forwards its decision to the  next terminal.  
	
	We characterize the relay and the receiver type-II error exponents  achieved by our schemes. Our  first scheme employs source coding without binning, which allows for a relatively simple  characterization of the achieved exponents. Our  second scheme employs source coding with binning and achieves larger exponents in some cases. However, with binning, the error exponent for $\hat{\mathcal{H}}_y$ is characterized by two competing exponents  and the exponent for $\hat{\mathcal{H}}_z$ by four competing exponents. They are thus more complicated to evaluate.
	
	In the  second part of the manuscript, we focus on two {cases}: the first is where $X^n \to Y^n \to Z^n$ forms a Markov chain under both hypotheses, and the second is where $X^n \to Z^n \to Y^n$ forms a Markov chain under both hypotheses. The first case models an extreme situation where  the relay lies in between the transmitter and the receiver, and thus the signals at the sensor and the receiver are conditionally independent given the signal at the relay. In such a situation, the two-hop network models, for example, short-range wireless communication where the sensor's signal only reaches the relay but not the more distant receiver. The second case models a situation 
 where the receiver lies in between the sensor and the relay, and thus the signals at the transmitter and the relay are conditionally independent given the signal at the receiver. In such a situation,  the two-hop network models, for example, communication in a cellular system where  the relay is a powerful {base station} and all communication goes through this {base station}. 
 
 We show that, in the first case where $X^n \to Y^n \to Z^n$, our schemes simplify considerably in the sense that the source coding scheme for the two-hop relay network  decouples into two independent point-to-point source coding schemes. In other words, it suffices to send quantization information about $X^n$ from the transmitter to the relay and, independently thereof,  to send quantization information about  $Y^n$ from the relay to the receiver (while also employing unanimous decision-forwarding) This contrasts the general scheme where the relay combines the quantization information about $X^n$ with its own observation $Y^n$ to create some kind of jointly processed quantization information to send to the receiver. The receiver error exponent achieved by  the simplified scheme equals the sum of the exponent at the relay and the exponent achieved over the point-to-point link from the relay to the receiver, but, to compute the second exponent, we modify the pmf of the relay's observation under $\mathcal{H}=1$ to being the same as its pmf under $\mathcal{H}=0$. These simplified expressions are proved to be optimal in different special cases of testing against independence (achieved without binning) and testing against conditional independence (with binning). The focus of this paper is on \emph{weak converses} where the type-I errors are also required to vanish asymptotically as $n\to \infty$. The existence of a \emph{strong converse} for one of these special cases, i.e., a proof that the same exponents are optimal also when type-I error probability  $\epsilon>0$ is tolerated,  was recently  proved in \cite{Tan}. 
	
	For the second case where $X^n\to Z^n \to Y^n$, we present optimality results (in the weak converse sense) for two special cases. In the first special case, $P_{YZ}$ is same under both hypothesis, so $Y^n$ by itself is of no interest to the receiver. For rates $T\ge R$, the optimal strategy is for the relay to ignore its own observation and simply forward the transmitter's message to the receiver. Interestingly, this simple forwarding strategy can become suboptimal when $T<R$, because then the relay can act as a ``coordinator'' to reduce the communication rate $T$ to the receiver. We present an example where the relay's own observation $Y^n$ allows the relay to extract the relevant portion of $X^n$, and thus to reduce the required rate to the receiver $T$. In the second special case, $P_{XZ}$ is same under both hypothesis, and for sufficiently large $T$ the optimal strategy for the relay is to ignore all communication from the transmitter. Again, using an example, we show that for small $T$ the transmitter can be useful by playing the role of a coordinator who reveals to the relay which portions of $Y^n$ are  relevant to the  receiver. 
	
	Lastly, as a side-result, we also present a detailed analysis of the Shimokawa-Han-Amari \cite{Amari} coding and testing scheme with binning for the point-to-point hypothesis testing problem. Previously this analysis has only appeared in Japanese \cite{Japanese}.

We conclude this introduction with remarks on notation and an outline of the paper.

\subsection{Notation}
We mostly follow the notation in \cite{ElGamal}.  Random variables are denoted by capital letters, e.g., $X,$ $Y,$ and their realizations by lower-case letters, e.g., $x,$ $y$.  Script symbols  such as $\mathcal{X}$ and $\mathcal{Y}$ stand for alphabets of  random variables, and $\mathcal{X}^n$ and $\mathcal{Y}^n$ for the corresponding $n$-fold Cartesian products.  Sequences of random variables $(X_i,...,X_j)$ and realizations $(x_i,\ldots, x_j)$ are  abbreviated by $X_i^j$ and $x_{i}^j$.  When $i=1$, then we also use $X^j$ and $x^j$ instead of $X_1^j$ and $x_{1}^j$. 

Generally, we write the probability mass function (pmf) of a discrete random variable $X$  as $P_X$; but we use $Q_X$ to indicate the pmf under hypothesis $\mathcal{H}=1$ when it is different from the pmf under $\mathcal{H}=0$. The conditional pmf of $X$ given $Y$  is written as  $P_{X|Y}$ (or as $Q_{X|Y}$ when $\mathcal{H}=1$). The distributions of $X^n$, $Y^n$ and $(X^n,Y^n)$ under the same hypothesis are denoted by $P_{X^n}$, $P_{Y^n}$ and $P_{X^nY^n}$, respectively. The notation $P_{XY}^n$ denotes the $n$-fold product distribution, i.e., 
 for every $(x^n,y^n)\in\mathcal{X}^n\times \mathcal{Y}^n$, we have:
\begin{align}
P_{XY}^n(x^n,y^n)=\prod_{i=1}^n P_{XY}(x_i,y_i).
\end{align}

The term   $D(P\| Q)$ stands for  the Kullback-Leibler (KL) divergence between two pmfs $P$ and $Q$ over the same alphabet.  
We use $\tp(\cdot)$ to denote the \emph{joint type} of a tuple.  For a joint type $\pi_{ABC}$ over alphabet $\mathcal{A}\times \mathcal{B}\times \mathcal{C}$, we denote by $I_{\pi_{ABC}}(A;B|C)$ the mutual information assuming that the random triple $(A,B,C)$ has  pmf  $\pi_{ABC}$; similarly for the entropy $H_{\pi_{ABC}}(A)$ and the conditional entropy $H_{\pi_{ABC}} (A|B)$. Sometimes we abbreviate $\pi_{ABC}$ by $\pi$. Also, when $\pi_{ABC}$ has been defined and is clear from the context, we write $\pi_{A}$ or $\pi_{AB}$ for the corresponding subtypes. When the type $\pi_{ABC}$ coincides with the actual pmf of a triple $(A,B,C)$, we  omit the subscript and simply write $H(A)$, $H(A|B)$, and $I(A;B|C)$.

For a given $P_X$ and a constant $\mu>0$, the set of sequences with the same type $P_X$ is denoted by $\mathcal{T}^n(P_X)$. We use  $\mathcal{T}_{\mu}^n(P_X)$ to denote the set of \emph{$\mu$-typical sequences} in $\mathcal{X}^n$:
\begin{align}
&\mathcal{T}_{\mu}^n(P_X)=\nonumber\\&\hspace{0.6cm}\Bigg\{x^n\colon \;\bigg| \frac{ | \{i\colon x_i=x\}|}{n}-P_X(x)\bigg|\leq \mu P_X(x), \;\;\forall x\in\mathcal{X}\Bigg\},\label{eq:typical}
\end{align}
where $|\{i\colon x_i=x\}|$ is the number of positions where the sequence $x^n$ equals  $x$.
Similarly,   $\mathcal{T}_{\mu}^n(P_{XY})$ stands for the set of \emph{jointly $\mu$-typical sequences} whose definition is as in \eqref{eq:typical} with $x$ replaced by $(x,y)$. 

The expectation operator is written as $\mathbb{E}[\cdot]$. The notation $\mathcal{U}\{a,\ldots, b\}$ is used to indicate a uniform distribution over the set $\{a,\ldots, b\}$; for the uniform distribution over $\{0,1\}$ we also use $\mathcal{B}(1/2)$. 
The $\log$ function is  taken with base $2$. Finally, we abbreviate left-hand side and right-hand side by LHS and RHS.

\subsection{Paper Outline} The remainder of the paper is organized as follows. 
	Section~\ref{seccascade} presents the problem description. Section~\ref{sec:no_binning}  presents a coding and testing scheme without binning and the exponent region it achieves.  Section~\ref{sec:binning} presents an improved scheme employing binning and the corresponding achievable exponent region. Sections~\ref{sec:MarkovXYZ}  and \ref{sec:MarkovXZY} study the proposed achievable regions when the Markov chains $X^n \to Y^n \to Z^n$ and $X^n \to Z^n \to Y^n$ hold, respectively.   The paper is concluded in Section~\ref{sec:conclusion} and by technical appendices.

\section{Detailed Problem Description}\label{seccascade}

Consider the multi-hop hypothesis testing problem with three terminals in Fig.~\ref{cascade}. The first terminal in the system, the \emph{transmitter},  observes the sequence $X^n$, the second terminal, the \emph{relay}, observes the sequence $Y^n$, and the third terminal, the \emph{receiver}, observes the sequence $Z^n$.  Under the null hypothesis 
\begin{align}
\mathcal{H}=0\colon \quad (X^n,Y^n,Z^n)\sim \text{i.i.d.}\; P_{XYZ},
\end{align}
whereas under the alternative hypothesis 
\begin{align}
\mathcal{H}=1\colon \quad (X^n,Y^n,Z^n)\sim \text{i.i.d.}\; Q_{XYZ},
\end{align}
for two given pmfs $P_{XYZ}$ and $Q_{XYZ}$. 

The problem encompasses a noise-free bit-pipe of rate $R$ from the transmitter to the relay and a noise-free bit pipe of rate $T$ from the relay to the receiver. That means, after observing $X^n$, the transmitter computes the message $M=\phi^{(n)}(X^n)$ using a possibly stochastic encoding function $\phi^{(n)}\colon \mathcal{X}^n\to \{0,...,\lfloor 2^{nR}\rfloor\}$ and sends it to the relay. The relay, after observing $Y^n$ and receiving $M$, computes the message   $B=\phi_y^{(n)}(M,Y^n)$ using  a possibly  stochastic encoding function 
$\phi_y^{(n)}\colon\mathcal{Y}^n\times \{0,...,\lfloor 2^{nR}\rfloor \}\to \{0,...,\lfloor 2^{nT}\rfloor\}$ and sends it to the receiver. 

The goal of the communication is that, based on their own observations and the received messages, the relay and the receiver can decide  on the hypothesis $\mathcal{H}$. 
The relay thus produces  the guess
\begin{align}
\hat{\mathcal{H}}_y=g_y^{(n)}(Y^n,M)
\end{align} 
using a guessing function $g_y^{(n)}\colon\mathcal{Y}^n\times \{0,....,\lfloor 2^{nR}\rfloor\}\to \{0,1\}$, 
and the receiver produces the guess
\begin{align}
\hat{\mathcal{H}}_z=g_z^{(n)}(Z^n,B)
\end{align}
using a guessing function $g_z^{(n)}\colon\mathcal{Z}^n\times \{0,...,\lfloor 2^{nT}\rfloor \}\to \{0,1\}$. 

\begin{definition}\label{cascadedef}
	For each $\epsilon \in (0,1)$, we say that the exponent-rate tuple $(\eta,\theta,R, T)$ is $\epsilon$-achievable if there exists a sequence of encoding and decoding functions $(\phi^{(n)},\phi_y^{(n)},g_y^{(n)},g_z^{(n)})$, $n=1,2,\ldots$, such that the corresponding sequences of type-I  and type-II error probabilities at the relay 
	\begin{align}
	\alpha_{y,n}&:{=} \Pr[\hat{\mathcal{H}}_y=1|\mathcal{H}=0],\\
	\beta_{y,n}&:{=} \Pr[\hat{\mathcal{H}}_y=0|\mathcal{H}=1],
	\end{align}
	and at the receiver
	\begin{align}
	\alpha_{z,n}&:{=} \Pr[\hat{\mathcal{H}}_z=1|\mathcal{H}=0],
	\\
	\beta_{z,n}&:{=} \Pr[\hat{\mathcal{H}}_z=0|\mathcal{H}=1],
	\end{align}
	satisfy 
	\begin{align}
	\alpha_{y,n}&\leq \epsilon,\label{def1b}\\
	\alpha_{z,n} & \leq \epsilon, \label{def1a}
	\end{align}
	and 
	\begin{align}
	-\varlimsup_{n\to \infty}\frac{1}{n}\log\beta_{y,n}&\geq \theta_y, \label{def2b}\\
	-\varlimsup_{n\to \infty}\frac{1}{n}\log\beta_{z,n}&\geq \theta_z.\label{def2a}
	\end{align} 
\end{definition}
\begin{definition}\label{def} For given rates $(R,T)$, we define the \emph{exponent-rate region $\mathcal{E}^*(R,T)$} as the closure of all non-negative  pairs $(\theta_y,\theta_z)$ for which $(\theta_y,\theta_z,R,T)$ is $\epsilon$-achievable for every $\epsilon\in(0,1)$. 
\end{definition}

\begin{remark}
In this paper we do not attempt to prove any ``strong converse.'' A strong converse in hypothesis testing would claim that the best achievable type-II error exponents for a given type-I error probability $\epsilon\in(0,1)$ does not depend on the value of $\epsilon$. For some special cases of the setting in Fig.~\ref{cascade}, a strong converse has recently been studied in \cite{Tan}.
\end{remark}

\section{A Coding  and Testing Scheme without Binning}\label{sec:no_binning}
 In this section we present  a first coding and testing scheme and characterize the achieved exponent-rate region using a relatively simple expression. The scheme is improved in the next section; the exponent-rate region achieved by the improved scheme is however more involved and includes multiple competing exponents.
 
 \subsection{The Coding and Testing Scheme} \label{sec:scheme}

Fix $\mu>0$, an arbitrary blocklength $n$, and  joint conditional pmfs $P_{SU|X}$ and $P_{V|SUY}$ over   finite auxiliary alphabets $\mathcal{S}$, $\mathcal{U}$, and $\mathcal{V}$. Define  the joint pmf 
\begin{equation}\label{eq:dist}
P_{SUVXYZ}=P_{XYZ}P_{SU|X}P_{V|SUY}
\end{equation} and the following nonnegative rates, which are calculated according to the  distribution in \eqref{eq:dist} and $\mu$:
\begin{align}
R_s&:{=}I(X;S)+\mu,\label{casnb1}\\
R_u&:{=}I(U;X|S)+\mu,\label{casnb2}\\
R_v&:{=}I(V;Y,U|S)+\mu.\label{casnb3}
\end{align}
Later, we shall choose the joint distributions in such a way that $R\ge R_s+R_u$ and $T\ge R_s+R_v$.

\underline{\textit{Code Construction}:} First, we randomly generate codewords 
\begin{equation}
\mathcal{C}_{S}:=
\left\{S^n(i)\colon \ i \in \{1,\ldots,\lfloor2^{nR_s}\rfloor\} \right\}
\end{equation}
by picking all entries i.i.d. according to $P_S$. Then, for each $i\in \{1,\ldots,\lfloor2^{nR_s}\rfloor\}$, we randomly generate codewords 
\begin{equation}
\mathcal{C}_{U}(i):=\left\{U^n(j|i)\colon\ j \in \{1,\ldots,\lfloor2^{nR_u}\rfloor\} \right\}
\end{equation} by choosing for each $t\in\{1,\ldots,n\}$ and $j \in \{1,\ldots,\lfloor2^{nR_u}\rfloor\}$, the $t$-th component  $U_t(j|i)$ of codeword $U^n(j|i)$ independently according to the conditional distribution $P_{U|S}(\cdot|S_t(i))$, where $S_t(i)$ denotes the $t$-th component of the codeword $S^n(i)$.  For each $i\in \{1,\ldots,\lfloor2^{nR_s}\rfloor\}$, generate also codewords
\begin{equation}
\mathcal{C}_{V}(i):=\left\{V^n(k|i)\colon k \in \{1,\ldots,\lfloor2^{nR_v}\rfloor\} \right\}
\end{equation} by choosing for each $t\in\{1,\ldots,n\}$ and $k \in \{1,\ldots,\lfloor2^{nR_v}\rfloor\}$ the $t$-th component $V_t(k|i)$ of codeword $V^n(k|i)$   independently according to the conditional distribution $P_{V|S}(\cdot|S_t(i))$.

Reveal the realizations $\{s^n(i)\}$, $\{u^n(j|i)\}$, and $\{v^n(k|i)\}$ of the random code constructions to all terminals.
\vspace{2mm}
	
\underline{\textit{Transmitter}:} Given that it observes the sequence $X^n=x^n$, the transmitter looks for a pair of indices $(i,j)$ such that
\begin{align}
(s^n(i),u^n(j|i),x^n)\in \mathcal{T}_{\mu/4}^{n}(P_{SUX}).
\end{align}
If successful, it picks one such pair uniformly at random and sends
\begin{equation}
M=(i,j)
\end{equation} to the relay. 
Otherwise, it sends $M=0$. 

\underline{\textit{Relay}:} Assume  that the relay observes the sequence $Y^n=y^n$ and receives the message $M=m$. If $m=0$, it declares $\hat{\mathcal{H}}_y=1$ and sends $b=0$ to the receiver. Otherwise, it decomposes  $m=(i,j)$ and
looks for  an index $k$ such that 
\begin{align}
(s^n(i),u^n(j|i),v^n(k|i),y^n)\in \mathcal{T}_{\mu/2}^{n}(P_{SUVY}).
\end{align} 
If such an index $k$ exists, 
the relay declares $\hat{\mathcal{H}}_y=1$ and sends the pair
\begin{equation}
B=(i,k)
\end{equation}
to the receiver. Otherwise, it declares $\hat{\mathcal{H}}_y=1$ and sends the message $B=0$.

\underline{\textit{Receiver}:} Assume that the receiver observes $Z^n=z^n$ and receives message $B=b$ from the relay. 
If $b=0$, the receiver declares $\hat{\mathcal{H}}_z=1$.
Otherwise, it decomposes $b=(i,k)$ and checks whether 
\begin{equation}
(s^n(i),v^n(k|i),z^n)\in \mathcal{T}_{\mu}^{n}(P_{SVZ}).
\end{equation}
If the typicality check is successful, the receiver declares $\hat{\mathcal{H}}_z=0$. Otherwise, it declares $\hat{\mathcal{H}}_z=1$. 
\vspace{3mm}

\subsection{Achievable Exponent-Rate Region}
	We present the exponent region achieved by the preceding scheme.

 Given two conditional pmfs $P_{SU|X}$ and $P_{V|SUY}$, define $\mathcal{E}_{\nobinning} (P_{SU|X}, P_{V|SUY})$ as the set of all  pairs $(\theta_y, \theta_z)$ that satisfy 
 	\begin{align}
 	\theta_y &\leq \min_{\substack{\tilde{P}_{SUXY}:\\\tilde{P}_{SUX}=P_{SUX}\\\tilde{P}_{SUY}=P_{SUY}}}D(\tilde{P}_{SUXY}\|P_{SU|X}Q_{XY}),\label{casach1}\\
 	\theta_z&\leq \min_{\substack{\tilde{P}_{SUVXYZ}:\\\tilde{P}_{SUX}=P_{SUX}\\\tilde{P}_{SUVY}=P_{SUVY}\\\tilde{P}_{SVZ}=P_{SVZ}}}D(\tilde{P}_{SUVXYZ}\|P_{SU|X}P_{V|SUY}Q_{XYZ}),\label{casach2}
 	\end{align}
 	where the joint pmf $P_{SUVXYZ}$ is defined as in \eqref{eq:dist} and $P_{SUX}$, $P_{SUY}$, $P_{SUVY}$ and $P_{SVZ}$ are marginals of this pmf.
 	
 	Define further the exponent region 
 	\begin{equation}\label{eq:Eidef}
\mathcal{E}_{\nobinning}(R,T) := \bigcup_{P_{SU|X}, P_{V|SUY} }\mathcal{E}_{\nobinning} (P_{SU|X}, P_{V|SUY})
  	\end{equation}
  	where the union is over all pairs of conditional pmfs $(P_{SU|X}, P_{V|SUY})$ satisfying
  	 	\begin{align}
  	 	R&\geq I(S,U;X),\label{casach3}\\
  	 	T&\geq I(X;S)+I(V;Y,U|S)\label{casach4}
  	 	\end{align}
  	 	and the  mutual informations are again calculated according to the joint pmf defined in \eqref{eq:dist}.
  	
\begin{theorem}[Achievable Region without Binning]\label{cascadeachnb}
For any pair of nonnegative rates $R,T \geq 0$, the set $\mathcal{E}_\textnormal{nobin}(R,T)$ is achievable:
\begin{equation}
\mathcal{E}_{\nobinning}(R,T) \subseteq \mathcal{E}^*(R,T)
\end{equation}
\end{theorem}
\begin{IEEEproof} See Appendix \ref{cascadeachnbpr}.\end{IEEEproof}

In the above theorem, it suffices to consider auxiliary random variables $S$, $U$, and $V$ over alphabets $\mathcal{S}$, $\mathcal{U}$, and $\mathcal{V}$ whose sizes satisfy:  $|\mathcal{S}| \leq |\mathcal{X}| + 4$, $|\mathcal{U}| \leq |\mathcal{X}|\cdot|\mathcal{S}|+3$ and $|\mathcal{V}| \leq |\mathcal{U}|\cdot|\mathcal{S}|\cdot |\mathcal{Y}|+2$. This follows by simple applications of Caratheodory's theorem.

\section{An Improved Scheme with Binning}\label{sec:binning}
In source coding, it is well known that binning can decrease the required rate of communication when observations at different terminals are correlated. The same holds for hypothesis testing. 
Before extending our coding and testing scheme from the previous section  to include binning, for completeness, we provide a detailed proof of the Shimokawa, Han, and Amari error exponent \cite{Amari} achieved over a point-to-point link when using binning.  So far, a detailed proof was available only in  Japanese \cite{Japanese}.
\subsection{Point-to-Point Link}
  
\begin{figure}[t!]
	\centering
	\includegraphics[scale=0.27]{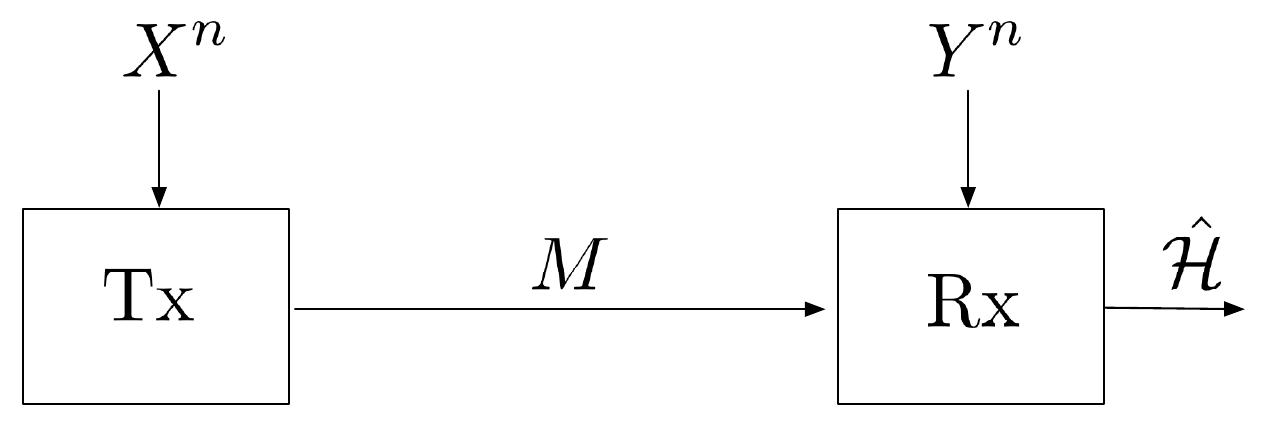}
	\caption{Hypothesis testing over a point-to-point channel}
	\label{p2p}
\end{figure}
Consider the network in Fig.~\ref{p2p}, which can be obtained as a special case from the previously introduced two-hop relay network by setting $T=0$ and $Z$ a constant that is the same under both hypotheses. In this case, the exponent  $\theta_z$ cannot be positive and is uninteresting. The system performance is then  characterized by the exponent $\theta_y$, and  for the purpose of this subsection, the relay can be regarded  as the final receiver. Therefore, in the remainder of this subsection, we call the terminal that observes $Y^n$ ``the receiver". We make the following definition: 
\begin{definition} Consider a single-hop system with only transmitter and receiver. 
The exponent-rate function $\theta^* (R)$ is the supremum of all $\epsilon$-achievable error exponents for a given rate $R$, i.e.,
	\begin{align}
	\theta^* (R) := \sup\ \{\theta_y \geq 0\colon (\theta_y, 0, R,0)\;\text{is}\;\epsilon\text{-achievable}\; \forall \epsilon>0\}.
	\end{align}
	\end{definition}

	We recall the lower bound  on $\theta^*(R)$ in \cite{Amari}, after presenting a coding and testing scheme that achieves this exponent. (The presented scheme slightly deviates  from the scheme in \cite{Amari}.)

\subsubsection{Coding and Testing Scheme}\label{sec:binning_scheme}

Fix $\mu>0$, a sufficiently large blocklength $n$, and the conditional pmf $P_{S|X}$ over a finite auxiliary alphabet $\mathcal{S}$. Define the joint pmf 
\begin{equation}\label{eq:PSXY}
P_{SXY}=P_{XY}P_{S|X}
\end{equation}
and, if $R> I(S;X)$ define  the nonnegative rate $R'=0$ and otherwise choose	 $R'$ such that
\begin{align}
R+R'& \geq  I(X;S)+\mu,\label{cas1binp2pb}\\
R' &< I(Y;S).\label{rateKLH} 
\end{align}
In the following coding scheme, when $R\leq I(S;X)$, we distribute the $s^n$-codewords in bins. Instead of sending the complete index of the chosen $s^n$, the transmitter sends only its bin number to the receiver. The receiver then selects the $s^n$ codeword from the indicated bin that is ``most-compatible" with its local observation $Y^n$, and makes its decision based on this selected codeword. By performing binning, the transmitter and the receiver can use a smaller communication rate, but the error probabilities may be higher.

\underline{\textit{Code Construction}:} Construct a random codebook
	\begin{align}
\mathcal{C}_{S}:=	\{S^n(m, \ell)\colon \ m \in\{1,...,\lfloor 2^{nR}\rfloor \},\ell\in\{1,...,\lfloor 2^{nR'}\rfloor \}\},\label{lembin0}
	\end{align}
	by drawing all entries of all codewords i.i.d. according to the chosen distribution $P_S$.
	
	Reveal the realization $\{s^n(m,\ell)\}$ of the random codebook to both terminals.
	\vspace{2mm}

\underline{\textit{Transmitter}:} Given that it observes the sequence $X^n=x^n$, the transmitter looks for indices $(m,\ell)$ such that 
\begin{align}\label{eq:typical_encoding}
(s^n(m,\ell),x^n)\in \mathcal{T}_{\mu/2}^n(P_{SX}).
\end{align} 
If successful, it picks one of these indices uniformly at random and sends the index $M=m$ to the relay. Otherwise, it sends $M=0$.
\vspace{2mm}

\underline{\textit{Receiver}:} Assume that the receiver observes $Y^n=y^n$ and receives the message $M=m$ from the transmitter. If $m=0$, the receiver declares $\hat{\mathcal{H}}=1$. Otherwise, it looks for an index $\ell'\in \{1,...,\lfloor 2^{nR'}\rfloor \}$ that minimizes $H_{\text{tp}(s^n(m, \ell''),y^n)}(S|Y)$ among all $\ell''$ satisfying $s^n(m,\ell'')\in\mathcal{T}_{\mu}^n(P_S)$.\footnote{Notice that because $m\neq 0$, there exists at least one codeword $s^n(m,\ell')\in \mathcal{T}_{\mu}^n(P_S)$ in bin $m$.} Then it checks whether 
$$(s^n(m,\ell'),y^n)\in \mathcal{T}_{\mu}^n(P_{SY}),$$ 
and  declares $\hat{\mathcal{H}}=0$ if this typicality check is successful and $\hat{\mathcal{H}}=1$ otherwise.
 \vspace{3mm}

\subsubsection{Result on the Error Exponent}
The scheme described in the previous subsection yields the following lower bound on the exponent-rate function. 

\vspace{2mm}

\begin{theorem}[\hspace{-0.1mm}{\cite{Amari}}]\label{thm_binning} For every choice of the conditional distribution $P_{S|X}$ satisfying that $R\ge I(S;X|Y)$, the exponent-rate function $\theta^*(R)$ is lower-bounded as 
	\begin{align}
	\theta^*(R) &\geq  \min \Bigg\{ \min_{\substack{\tilde{P}_{SXY}:\\\tilde{P}_{SX}=P_{SX}\\\tilde{P}_{SY}=P_{SY}}}\hspace{-0.2cm} D(\tilde{P}_{SXY}\|P_{S|X}Q_{XY}), \nonumber \\
	& \hspace{-.3cm} \min_{\substack{\tilde{P}_{SXY}:\\\tilde{P}_{SX}=P_{SX}\\\tilde{P}_Y=P_Y\\H(S|Y) \leq H_{\tilde{P}_{SY}}(S|Y)}}\hspace{-0.9cm} D(\tilde{P}_{SXY}\|P_{S|X}Q_{XY})+ R-I(S;X|Y)\Bigg\}.\label{ach}
	\end{align}	
For a choice of $P_{S|X}$ such that $R\geq  I(S;X)$, the bound can be tightened to
	\begin{align}
	\theta^*(R) &\geq  \min_{\substack{\tilde{P}_{SXY}:\\\tilde{P}_{SX}=P_{SX}\\\tilde{P}_{SY}=P_{SY}}} D(\tilde{P}_{SXY}\|P_{S|X}Q_{XY}),\label{eq:LB}
	\end{align}	
Here   mutual informations 
 and the entropy $H(S|Y)$ in the miniminization constraint are calculated according to the joint pmf in \eqref{eq:PSXY} and the chosen conditional pmf $P_{S|X}$.
\end{theorem}
\begin{IEEEproof}When $R\geq I(S;X)$,  our scheme does not use binning and an analysis similar to Appendix~\ref{cascadeachnbpr} (the analysis of the multi-hop scheme without binning) yields the desired result. When $R<I(S;X)$, our scheme uses binning and is analyzed in Appendix \ref{thm1pr}.\end{IEEEproof}

In the above theorem, it suffices to consider  auxiliary random variables $S$ over alphabets $\mathcal{S}$ whose sizes satisfy:  $|\mathcal{S}| \leq |\mathcal{X}| + 2$.


The inequality in Theorem \ref{thm_binning} holds with equality in  the  special cases of \emph{testing against independence} \cite{Csiszar86}, where  $Q_{XY}=P_X \cdot P_Y$,\footnote{There is no need to apply the coding scheme with binning to attain the optimal error exponent in this case, see \cite{Csiszar86}.} and of \emph{testing against conditional independence} \cite{Wagner}, where $Y$ decomposes as $Y=(\YC,\YH)$ and $Q_{X\YC\YH}=P_{X\YC} P_{\YH|\YC}$.

\subsection{The Two-Hop Relay Network}\label{sec:cascade_binning}

We turn back to the two-hop relay network and propose an improved coding and testing scheme employing binning. 

\subsubsection{Coding and Testing Scheme}

Fix $\mu>0$, an arbitrary blocklength $n$, and  joint conditional pmfs $P_{SU|X}$ and $P_{V|SUY}$ over   finite auxiliary alphabets $\mathcal{S}$, $\mathcal{U}$, and $\mathcal{V}$. Define  the joint pmf $P_{SUVXYZ}=P_{XYZ}P_{SU|X}P_{V|SUY}$ and the following nonnegative rates, which are calculated according to the chosen distribution,
\begin{subequations}\label{eq:rates_cascd}
\begin{align}
R_s&= I(X;S)+\mu,\label{cas1bin}\\
R_u+R'_u&= I(U;X|S)+\mu,\label{cas2bin}\\
R_v+R'_v&= I(V;Y,U|S)+\mu,\label{cas3bin}\\
R'_u&\leq I(U;Y|S),\label{cas5bin}\\
R'_v&\leq I(V;Z|S).\label{cas6bin}
\end{align}
\end{subequations}
The joint distributions are chosen  in such a way that 
\begin{IEEEeqnarray}{rCl}
	R& \ge &R_s+R_u \\ 
	T & \ge  & R_s+R_v.
	\end{IEEEeqnarray}

In the following coding scheme, the transmitter distributes the $u^n$ codewords in bins, and sends the bin number of the chosen $u^n$ to the relay. The relay looks in that bin for the  $u^n$ codeword that is ``most compatible" with its local observation $Y^n$. Similarly, the relay and the receiver perform binning on $v^n$. Note that for simplicity and ease of exposition, we do not bin the $s^n$ codewords.

\underline{\textit{Code Construction}:} Construct a random codebook $$\mathcal{C}_{S}= \{S^n(i) \colon  \;\;   i\in\{1, ..., \lfloor 2^{nR_s}\rfloor\} \}$$
by selecting each entry of the $n$-length codeword $s^n(i)$ in an i.i.d. manner according to the pmf $P_{S}$. Then, for each $i$,  generate random codebooks 
\begin{IEEEeqnarray*}{rCl}
\lefteqn{\mathcal{C}_{U}(i)=}~~\nonumber\\*&&\{U^n(j,e|i)\colon  j\in \{1,...,\lfloor 2^{nR_u}\rfloor\}, e\in \{1,...,\lfloor 2^{nR'_u}\rfloor\}\}
\end{IEEEeqnarray*}
and 
\begin{IEEEeqnarray*}{rCl}
\lefteqn{\mathcal{C}_{V}(i)=}~~~\nonumber\\*
&&\{V^n(k,f|i)\colon  k\in \{1,...,\lfloor 2^{nR_v}\rfloor \}, f\in \{1,...,\lfloor 2^{nR'_v}\rfloor \}\}
\end{IEEEeqnarray*}
 by selecting for each $t\in\{1,\ldots, n\}$, the $t$-th components $U_t(j,e|i)$ and $V_t(k,f|i)$  of  the  codewords $U^n(j,e|i)$ and $V^n(k,f|i)$ independently using the conditional pmfs $P_{U|S}(\cdot|S_t(i))$ and $P_{V|S}(\cdot|S_t(i))$, where $S_t(i)$ denotes the $t$-th component of codeword $S^n(i)$. 
 
 Reveal the realizations $\{s^n(i)\}$, $\{u^n(j,e|i)\}$, and $\{v^n(k,f|i)\}$ of the random codebooks to all terminals.

\vspace{2mm}

\underline{\textit{Transmitter}:} Given that the transmitter observes the sequence $X^n=x^n$, it looks for indices $(i,j,e)$ such that 
\begin{align}\label{eq:typical_cascade_encoding}
(s^n(i),u^n(j,e|i),x^n)\in \mathcal{T}_{\mu/4}^n(P_{SUX}).
\end{align} 
If successful, it picks one such triple uniformly at random, and sends the first two indices of the triple:
\begin{align} 
M= (i,j)
\end{align}
to the relay. Otherwise, it sends $M=0$.
\vspace{2mm}

\underline{\textit{Relay}:} Assume that the relay observes the sequence $Y^n=y^n$ and receives the message $M=m$. If $m=0$, it declares $\hat{\mathcal{H}}_y=1$ and sends $B=0$ to the receiver. Otherwise, it looks for an index $e'$ which minimizes $H_{\text{tp}(s^n(i),u^n(j,e''|i),y^n)}(U|S,Y)$ among all $e''$ satisfying $(s^n(i),u^n(j,e''|i))\in \mathcal{T}_{\mu/2}^n(P_{SU})$.
It then looks for  indices $(k,f)$ such that 
\begin{align}
(s^n(i),u^n(j,e'|i),v^n(k,f|i),y^n)\in \mathcal{T}_{\mu/2}^n(P_{SUVY}).
\end{align}
If successful, it declares $\hat{\mathcal{H}}_y=0$ and picks one of these index pairs uniformly at random. It then sends the corresponding indices
\begin{align}
B=(i,k)
\end{align}
 to the receiver. Otherwise, it declares $\hat{\mathcal{H}}_y=1$ and sends the message $B=0$ to the receiver. 
\vspace{2mm}

\underline{\textit{Receiver}:} Assume that the receiver observes $Z^n=z^n$ and receives message $B=b$ from the relay. If $b=0$, the receiver declares $\hat{\mathcal{H}}_z=1$. Otherwise, it looks for an index $f'$ which minimizes $H_{\text{tp}(s^n(i),v^n(k,f''|i),z^n)}(V|S,Z)$ among all $f''$ satisfying $(s^n(i),v^n(k,f''|i))\in \mathcal{T}_{\mu/2}^n(P_{SV})$.
Then, it checks whether 
\begin{align}
(s^n(i),v^n(k,f'|i),z^n)\in \mathcal{T}_{\mu}^n (P_{SVZ}).
\end{align}
If successful, the receiver declares $\hat{\mathcal{H}}_z=0$. Otherwise, it declares $\hat{\mathcal{H}}_z=1$.
\vspace{2mm}

\subsubsection{Result on the Exponent-Rate  Region}
The coding scheme in the previous subsection establishes the following theorem. 

For any pair of conditional pmfs $P_{US|X}$ and $P_{V|SUY}$, let  $\mathcal{E}_{\binning}(P_{US|X}, P_{V|SUY})$ denote the set of all exponent-pairs $(\theta_y,\theta_z)$ that satisfy
		\begin{align}
		\theta_y &\leq \min \{ \theta_y^{(1)},\theta_y^{(2)}\},\label{casach1bin}\\
		\theta_z&\leq \min\{\theta_z^{(i)}\colon i=1,...,4\},\label{casach2bin}
		\end{align}
where 
\begin{subequations}\label{eq:theta10}
	\begin{IEEEeqnarray}{rCl}
		\theta_y^{(1)} &:=& \min_{\substack{\tilde{P}_{SUXY}:\\\tilde{P}_{SUX}=P_{SUX}\\\tilde{P}_{SUY}=P_{SUY}}} D(\tilde{P}_{SUXY}\|P_{SU|X}Q_{XY}),\\[1ex]
	\theta_y^{(2)}  &:=&\hspace{-0.1cm} \min_{\substack{\tilde{P}_{SUXY}:\\\tilde{P}_{SUX}=P_{SUX}\\\tilde{P}_{SY}=P_{SY}\\H(U|S,Y)\leq H_{\tilde{P}}(U|S,Y)}}\!\!\!\!  D(\tilde{P}_{SUXY}\|P_{SU|X}Q_{XY})\nonumber\\&&\hspace{1.5cm}{}+R-I(S,U;X)+I(U;Y|S),	\IEEEeqnarraynumspace \\[1ex]
	\theta_z^{(1)}  &:=& \min_{\substack{\tilde{P}_{SUVXYZ}:\\\tilde{P}_{SUX}=P_{SUX}\\\tilde{P}_{SUVY}=P_{SUVY}\\\tilde{P}_{SVZ}=P_{SVZ}}}\!\!\!\!  D(\tilde{P}_{SUVXYZ}\|P_{SU|X}P_{V|SUY}Q_{XYZ}),\nonumber\\\label{th1last}\\[2ex]
		\theta_z^{(2)}  &:=&\hspace{-0.3cm} \min_{\substack{\tilde{P}_{SUVXYZ}:\\\tilde{P}_{SUX}=P_{SUX}\\\tilde{P}_{SVY}=P_{SVY}\\\tilde{P}_{SVZ}=P_{SVZ}\\H(U|S,Y)\leq H_{\tilde{P}}(U|S,Y)}}\!\!\!\!\!\!\!\!\!\! D(\tilde{P}_{SUVXYZ}\| P_{SU|X}P_{V|SY}Q_{XYZ})\nonumber\\&&\hspace{1.5cm}{}+R-I(S,U;X) + I(U;Y|S), \\[1ex]
	\theta_z^{(3)}  &:=& \hspace{-3mm}\min_{\substack{\tilde{P}_{SUVXYZ}:\\\tilde{P}_{SUX}=P_{SUX}\\\tilde{P}_{SUVY}=P_{SUVY}\\\tilde{P}_{SZ}=P_{SZ}\\H(V|S,Z)\leq H_{\tilde{P}}(V|S,Z)}} \hspace{-0.5cm} D(\tilde{P}_{SUVXYZ}\|P_{SU|X}P_{V|SUY}Q_{XYZ})\nonumber\\&&\hspace{0.2cm}{}+T-I(S;X)-I(V;Y,U|S)+I(V;Z|S),\\[1ex]
	\theta_z^{(4)} &:=&\hspace{-0.3cm} \min_{\substack{\tilde{P}_{SUVXYZ}:\\\tilde{P}_{SUX}=P_{SUX}\\\tilde{P}_{SVY}=P_{SVY}\\\tilde{P}_{SZ}=P_{SZ}\\H(U|S,Y)\leq H_{\tilde{P}}(U|S,Y)\\H(V|S,Z)\leq H_{\tilde{P}}(V|S,Z)}}\!\!\!\! D(\tilde{P}_{SUVXYZ}\| P_{SU|X}P_{V|SY}Q_{XYZ}) \nonumber \\
	& & \hspace{0.3cm}+ R+T-I(S,U;X)-I(X;S)-I(V;U,Y|S)\nonumber\\&&\hspace{2cm}{}+I(U;Y|S)+I(V;Z|S),
	\end{IEEEeqnarray}
\end{subequations}
where the mutual information and entropy terms, as well as  the marginals $P_{SUX}$, $P_{SUVY}$, $P_{SVY}$,  $P_{SVZ}$, and $P_{SZ}$ are calculated with respect to the joint pmf
\begin{equation}\label{eq:joint}
P_{SUVXYZ} = P_{US|X}P_{V|USY} P_{XYZ}.
\end{equation}
Define then the exponent-rate region
\begin{equation}
 \mathcal{E}_{\binning}(R,T):=
 \bigcup_{(P_{US|X},P_{V|USY})}   \mathcal{E}_{\binning}(P_{US|X},P_{V|USY})
 \end{equation}
 where the union is over all pairs of conditional distributions so that 
 the rate constraints  
\begin{align}
R&\geq I(S,U;X)-I(U;Y|S),\label{casach5bin}\\
T&\geq I(V;Y,U|S)-I(V;Z|S)+I(S;X),\label{casach7bin}
\end{align}
are satisfied when the mutual informations are calculated according to the joint pmf in \eqref{eq:joint}.

\begin{theorem}[Achievable Region with Binning]\label{thm:casbin} For any positive rate-pair $(R,T)$:
	\begin{equation}
	\mathcal{E}_{\binning}(R,T) \subseteq 	\mathcal{E}^*(R,T).
	\end{equation}
\end{theorem}

\begin{IEEEproof} See Appendix \ref{casbinpr}.\end{IEEEproof}
\medskip

For each choice of conditional pmfs $P_{US|X}$ and $P_{V|USY}$, the achievable exponents-region $\mathcal{E}_{\textnormal{bin}}(P_{US|X}, P_{V|USY})$ is characterized through two competing  exponents at the relay and four competing  exponents at the receiver, see \eqref{casach1bin} and \eqref{casach2bin}. Extending 
 	our scheme by
binning also the $s^n$ codewords achieves an  exponents region that  is characterized by three competing  exponents at the relay and ten competing exponents at the receiver.	Details are omitted for brevity.

\section{The Special Case ``$X^n \to Y^n \to Z^n$ Under Both Hypotheses"}\label{sec:MarkovXYZ}
Consider a situation  where  the relay lies in between the transmitter and the receiver, and thus the signals at the sensor and the receiver are conditionally independent given the signal at the relay. In this situation, the two-hop relay network seems particularly adequate for modelling short-range wireless communication.

Assume that the pmfs $P_{XYZ}$ and $Q_{XYZ}$ decompose as
\begin{align}
P_{XYZ} &= P_X\cdot P_{Y|X}\cdot P_{Z|Y},\label{case0a}\\
Q_{XYZ} &= Q_X\cdot Q_{Y|X}\cdot Q_{Z|Y}.\label{case0b}
\end{align}

We start by showing that in this special case the compression mechanisms in the previously-presented coding and testing schemes can be simplified. There is no need to send compression information from the transmitter to the receiver. Hence, the message sent from the relay to the receiver consists only of the relay's own guess and compression information of the relay's observation. 
Technically, this means that the expressions for  $\mathcal{E}_\nobinning(R,T)$ and $\mathcal{E}_\binning(R,T)$ can be simplified for this special case by setting $S$ to be a constant, and choosing $V$ to be conditionally independent of $U$ given $Y$. In the following, we use the subscript ``dcpled'' to refer to the region of this special case, which stands for ``decoupled''. Here, the transmitter-relay and relay-receiver links are basically decoupled from each other thanks to the Markov chain $X\to Y\to Z$.

\subsection{Simplified Exponent Regions} 
Given two conditional pmfs $P_{U|X}$ and $P_{V|Y}$, define the exponents region
 $\mathcal{E}_{\simple}(P_{U|X}, P_{V|Y})$ as the set of all  pairs $(\theta_y, \theta_z)$ that satisfy 
\begin{align}
\theta_y &\leq \min_{\substack{\tilde{P}_{UXY}:\\\tilde{P}_{UX}=P_{UX}\\\tilde{P}_{UY}=P_{UY}}}D(\tilde{P}_{UXY}\|P_{U|X}Q_{XY}),\\
\theta_z&\leq\min_{\substack{\tilde{P}_{UXY}:\\\tilde{P}_{UX}=P_{UX}\\\tilde{P}_{UY}=P_{UY}}}D(\tilde{P}_{UXY}\|P_{U|X}Q_{XY}) \nonumber\\&\hspace{1cm}+ \min_{\substack{\tilde{P}_{VYZ}:\\\tilde{P}_{VY}=P_{VY}\\\tilde{P}_{VZ}=P_{VZ}}}  \mathbb{E}_{P_{Y}}\Big[  D(\tilde{P}_{VZ|Y}\|P_{V|Y}Q_{Z|Y})\Big],
\end{align}
where $P_{UY}$ and $P_{VZ}$ indicate the marginals of the joint pmfs $P_{U|X}P_{XY}$ and $P_{V|Y} P_{YZ}$, and further define 
\begin{equation}
\mathcal{E}_\simple(R,T) := \bigcup_{P_{U|X}, P_{V|Y} }\mathcal{E}_{\simple} (P_{U|X}, P_{V|Y})
\end{equation}
where the union is over all pairs of conditional pmfs $(P_{U|X}, P_{V|Y})$ satisfying
\begin{align}
R&\geq I(U;X),\label{casach3b}\\
T&\geq I(V;Y)\label{casach4b}
\end{align}
for  mutual informations that are calculated according to the joint pmfs $P_{UX}=P_{U|X}P_X$ and $P_{VY}=P_{V|Y}P_Y$.  

\begin{proposition}[Simplified Achievable Region Without Binnning]\label{prop:Markov}
	If \eqref{case0a} and \eqref{case0b} hold, 
	then 
	\begin{equation}
	\mathcal{E}_\simple (R,T) = \mathcal{E}_{\nobinning}(R,T).
	\end{equation}
\end{proposition}
\begin{IEEEproof}
	See Appendix~\ref{app:proof_prop}.
\end{IEEEproof}

In the above proposition, it suffices to consider auxiliary random variables $U$ and $V$ over alphabets $\mathcal{U}$ and $\mathcal{V}$ whose sizes satisfy:  $|\mathcal{U}| \leq |\mathcal{X}|+1$ and $|\mathcal{V}| \leq  |\mathcal{Y}|+1$.

\bigskip

Similarly, given two conditional pmfs $P_{U|X}$ and $P_{V|Y}$, let  $\mathcal{E}_{\binsimple}(P_{U|X}, P_{V|Y})$ denote the set of all exponent-pairs $(\theta_y,\theta_z)$ that satisfy
		\begin{subequations}
	\begin{IEEEeqnarray}{rCl}
		\theta_y &\leq& \min \Big\{ \min_{\substack{\tilde{P}_{UXY}:\\\tilde{P}_{UX}=P_{UX}\\\tilde{P}_{UY}=P_{UY}}} D(\tilde{P}_{UXY}\|P_{U|X}Q_{XY}),\nonumber\\&&\hspace{-0.3cm} \min_{\substack{\tilde{P}_{UXY}:\\\tilde{P}_{UX}=P_{UX}\\\tilde{P}_{Y}=P_{Y}\\H(U|Y)\leq H_{\tilde{P}}(U|Y)}}\hspace{-0.6cm}  D(\tilde{P}_{UXY}\|P_{U|X}Q_{XY})+R-I(U;X|Y)\Big\}, \nonumber\\
		\end{IEEEeqnarray}
		and 
		\begin{IEEEeqnarray}{rCl}
	\theta_z & \leq & \min \Big\{ \min_{\substack{\tilde{P}_{UXY}:\\\tilde{P}_{UX}=P_{UX}\\\tilde{P}_{UY}=P_{UY}}} D(\tilde{P}_{UXY}\|P_{U|X}Q_{XY}),\nonumber\\&&\hspace{0.6cm} \min_{\substack{\tilde{P}_{UXY}:\\\tilde{P}_{UX}=P_{UX}\\\tilde{P}_{Y}=P_{Y}\\H(U|Y)\leq H_{\tilde{P}}(U|Y)}} \hspace{-0.4cm}D(\tilde{P}_{UXY}\|P_{U|X}Q_{XY})\nonumber\\[-3ex]&&\hspace{4cm} +R-I(U;X|Y)\Big\},  \nonumber \\[2ex]
	& & +  \min\Big\{  \min_{\substack{\tilde{P}_{VZ|Y}:\\\tilde{P}_{VY}=P_{VY}\\\tilde{P}_{VZ}=P_{VZ}}}  \mathbb{E}_{P_{Y}}\Big[  D(\tilde{P}_{VZ|Y}\|P_{V|Y}Q_{Z|Y})\Big],\nonumber\\&&\hspace{0.7cm} \min_{\substack{\tilde{P}_{VZ|Y}:\\\tilde{P}_{VY}=P_{VY}\\\tilde{P}_{Z}=P_{Z}\\H(V|Z)\leq H_{\tilde{P}}(V|Z)}} \hspace{-0.2cm}  \mathbb{E}_{P_{Y}}\Big[  D(\tilde{P}_{VZ|Y}\|P_{V|Y}Q_{Z|Y})\Big]\nonumber\\[-3ex]&&\hspace{4cm}+T-I(V;Y|Z)\Big\},\nonumber\\\label{eq:simpletheta_zbinning}
	\end{IEEEeqnarray}
\end{subequations}
 where the mutual information and entropy terms, as well as  the marginals $P_{SUX}$, $P_{SUVY}$, $P_{SVY}$,  $P_{SVZ}$, and $P_{SZ}$ are calculated with respect to the joint pmf
\begin{equation}\label{eq:joint2}
P_{SUVXYZ} = P_{US|X}P_{V|USY} P_{XYZ}.
\end{equation}
Further define
\begin{equation}
 \mathcal{E}_{\binsimple}(R,T):=
 \bigcup_{{P_{U|X},P_{V|Y}}}   \mathcal{E}_{\binsimple}(P_{U|X},P_{V|Y})
 \end{equation}
 where the union is over all pairs of conditional distributions for which
 the rate constraints  
\begin{align}
R&\geq I(U;X|Y),\label{casach5bin}\\
T&\geq I(V;Y|Z),\label{casach7bin}
\end{align}
are satisfied when the mutual informations are calculated according to the joint pmf in \eqref{eq:joint2}.

\begin{proposition}[Simplified Achievable Region With Binnning]\label{cor-bin-state} 
	If \eqref{case0a} and \eqref{case0b} hold, 
	then 
	\begin{equation}
	\mathcal{E}_ \binsimple (R,T) = \mathcal{E}_{\binning}(R,T).
	\end{equation}
\end{proposition}
\begin{IEEEproof}
The inclusion $\mathcal{E}_ \binsimple (R,T) \subseteq \mathcal{E}_{\binning}(R,T)$ follows by restricting to $U$ and $V$ to be conditionally independent given $Y$ and $S$ to be a constant. The proof of inclusion 
$\mathcal{E}_{\binsimple} (R,T) \supseteq \mathcal{E}_{\binning}(R,T)$ is sketched in Appendix~\ref{app:inclusion2}. 
\end{IEEEproof}

\medskip

\begin{remark} For both Propositions~\ref{prop:Markov} and \ref{cor-bin-state}, the exponent at the receiver equals the sum of two exponents: the first is the exponent at the relay (i.e., the exponent attained over the transmitter-relay link), and the second is the exponent on the isolated relay-receiver link, but with $Q_{YZ}$ replaced by $P_Y Q_{Z|Y}$. 
\end{remark}

\subsection{Optimality Results}
In the following, we prove optimality of the achievable region in Proposition~\ref{cor-bin-state} for some cases of ``testing against conditional independence'' under the Markov conditions \eqref{case0a} and \eqref{case0b}. In the following examples, if the random variables $\YC$ and $\ZC$ are constants, then the setups reduce to ``testing against independence". For ``testing against independence", achievability can also be established using  the simpler Proposition~\ref{prop:Markov}. In other words,  the optimal exponents can also be achieved without binning.
\bigskip

\subsubsection{Special Case 1}
Assume that the relay's and the receiver's observations decompose as 
\begin{IEEEeqnarray}{rCl}
	Y & = &(\YC,\YH) \label{SI1}\\
	Z & = & (\YC,\ZC,\ZH)
	\end{IEEEeqnarray}
	and 
	\begin{align}
\textnormal{under } \mathcal{H}=0 &\colon \;\; \big(X^n,\YC^{n},\YH^n,\ZC^n, \ZH^n\big)\;\text{i.i.d.}\nonumber\\&\hspace{1cm} \sim  P_{X|\YC\YH}\cdot P_{\YC\YH\ZC\ZH},\label{SI2b}\\
\textnormal{under }\mathcal{H}=1 &\colon \;\;  \big(X^n,\YC^{n},\YH^n,\ZC^n, \ZH^n\big)\;\text{i.i.d.}\nonumber\\&\hspace{1cm} \sim P_{X|\YC}\cdot P_{\YH|\YC\ZC}\cdot P_{\YC\ZC\ZH}.\label{SI2}
\end{align}
 The  following corollary shows that in this case, the receiver's optimal error exponent  equals the \emph{sum} of the optimal error exponent at the relay and the optimal error exponent achieved over the isolated relay-receiver link.
 
\begin{corollary}\label{SIthm} If  \eqref{SI1}--\eqref{SI2} hold, the exponent-rate  region $\mathcal{E}^*(R,T)$ is the set of all nonnegative pairs $(\theta_y,\theta_z)$ that satisfy
	\begin{align}
	\theta_y &\leq I(U;Y|\YC),\\
	\theta_z &\leq I(U;Y|\YC)+I(V;Z|\ZC,\YC),\label{thbinopt}
	\end{align}
		for some auxiliary random variables $(U,V)$ satisfying the Markov chains $U\to X\to Y$ and $V\to Y\to Z$ and the rate constraints
	\begin{align}
	R &\geq I(U;X|\YC),\\
	T &\geq I(V;Y|\YC,\ZC),
	\end{align}
and where $Y=(\YC,\YH)$, $Z=(\ZC,\ZH)$, and $(X,\YC,\YH,\ZC, \ZH) \sim  P_{X|\YC\YH}\cdot P_{\YC\YH\ZC\ZH}$.
	\end{corollary}
\begin{IEEEproof} Achievability follows by simplifying Proposition~\ref{cor-bin-state}. For  this special case, since $R \geq I(U;X|\YC)$ and $T\geq I(V;Y|\YC,\ZC)$, exponents $\theta_{z}^{(2)},\theta_{z}^{(3)},\theta_{z}^{(4)}$ become inactive in view of $\theta_{z}^{(1)}$. The converse is proved in Appendix~\ref{SIproof}.
	\end{IEEEproof}

In the above theorem it suffices to consider auxiliary random variables $U$ and $V$ over alphabets  $\mathcal{U}$ and $\mathcal{V}$ whose sizes satisfy:  $|\mathcal{U}| \leq |\mathcal{X}| + 2$ and $|\mathcal{V}| \leq |\mathcal{Y}|+1$.

\begin{remark}
If we set $\YC$ and $\ZC$ to constants, then this special case reduces to one where
\begin{IEEEeqnarray}{rCl}
P_{XYZ} & = & P_{XY}\cdot P_{Z|Y}\\
Q_{XYZ} & = & P_X\cdot P_{Y} \cdot P_Z.
\end{IEEEeqnarray}
The exponent-region then becomes the set of all nonnegative pairs $(\theta_y, \theta_z)$  that satisfy
	\begin{align}
	\theta_y &\leq I(U;Y),\label{thm2aa}\\
	\theta_z &\leq I(U;Y)+I(V;Z),\label{thm2bb}
	\end{align}
	for a pair of auxiliary random variables $U$ and $V$ satisfying the Markov chains  $U\to X\to Y$ and $V\to Y \to Z$  and the rate constraints
	\begin{align}
	R &\geq I(U;X)\label{thm2cc}\\
		T &\geq I(V;Y).\label{thm2d}
	\end{align}
Furthermore, the exponent-rate region can be obtained using Proposition~\ref{prop:Markov}.
\end{remark}

\bigskip

\subsubsection{Special Case 2}

Assume that the receiver's observation decomposes as 
\begin{IEEEeqnarray}{rCl} \label{eq:decomp1}
	Z & = & (\ZC,\ZH)
	\end{IEEEeqnarray}
	and 
	\begin{align}
\textnormal{under } \mathcal{H}=0 &\colon \quad  \big(X^n,Y^n,\ZC^n, \ZH^n\big)\;\text{i.i.d.}\; \sim  P_{XY\ZC\ZH}, \label{eq:hyp1}\\
\textnormal{under }\mathcal{H}=1 &\colon \quad  \big(X^n, Y^n,\ZC^n, \ZH^n\big)\;\text{i.i.d.}\; \sim P_{X Y\ZC}\cdot  P_{\ZH|\ZC}.\label{eq:hyp2}
\end{align}
In this case, the relay cannot obtain a positive exponent since $(X^n,Y^n)\sim P_{XY}$  under both hypotheses. Moreover, as the following corollary shows, the relay can completely ignore the message from the transmitter and act as if it was the transmitter of a simple point-to-point setup \cite{Han}.
\begin{corollary}\label{prop1b} Assume \eqref{eq:decomp1}--\eqref{eq:hyp2}. 	
	The exponent-rate region $\mathcal{E}^*(R,T)$ is the set of all nonnegative pairs $(\theta_y, \theta_z)$  that satisfy
	\begin{align}
	\theta_y &=0\\
	\theta_z &\leq I(V;\ZH|\ZC)\label{thm:oneb}
	\end{align}
	for an auxiliary random variable $V$ satisfying the Markov chain  $V\to Y \to Z$  and the rate constraint
	\begin{align}
	T &\geq I(V;Y|\ZC),
	\end{align}
where $Z=(\ZC,\ZH)$, and $(X,Y,\ZC, \ZH) \sim  P_{XY\ZC\ZH}$.
		(No constraint involves the rate $R$.)
\end{corollary}
\begin{IEEEproof} Achievability follows by specializing Proposition~\ref{cor-bin-state} to  $U = 0$ (deterministically) and then simplifying the expressions. In particular, notice that, since $T\geq I(V;Y |\ZC)$, exponents  $\theta_{z}^{(2)},\theta_{z}^{(3)},\theta_{z}^{(4)}$ become inactive in view of $\theta_{z}^{(1)}$. The converse is standard; details can be found in Appendix~\ref{app:convCor}.
\end{IEEEproof}

\begin{remark}
If we set $\ZC$ to a constant, then the problem reduces to one where 
\begin{IEEEeqnarray}{rCl}
P_{XYZ} & = & P_{XY} \cdot P_{Z|Y}\\
Q_{XYZ} & = & P_{XY} \cdot P_Z.
\end{IEEEeqnarray}
The exponent-rate region then becomes the set of all nonnegative pairs $(\theta_y, \theta_z)$  that satisfy
	\begin{align}
	\theta_y &=0,\\
	\theta_z &\leq I(V;Z),
	\end{align}
	for an auxiliary random variable $V$ satisfying the Markov chain  $V\to Y \to Z$  and the rate constraint
	\begin{align}
	T &\geq I(V;Y).
	\end{align}
The region is again achievable using Proposition~\ref{prop:Markov}.
\end{remark}

\bigskip

\subsubsection{Special Case 3}

Assume that the relay's  observation decomposes as 
\begin{IEEEeqnarray}{rCl}
	Y & = &(\YC,\YH), \label{eq:set1}
		\end{IEEEeqnarray}
	and 
	\begin{align}
\textnormal{under } \mathcal{H}=0 &\colon \quad  \big(X^n,\YC^{n},\YH^n,Z^n\big)\;\text{i.i.d.}\; \sim  P_{X\YC\YH Z},\\
\textnormal{under }\mathcal{H}=1 &\colon \quad  \big(X^n,\YC^{n},\YH^n,Z^n\big)\;\text{i.i.d.}\; \sim P_{X|\YC}\cdot P_{\YC\YH Z}.\label{eq:set3}
\end{align}
\medskip

As the following corollary shows, in this case the  optimal strategy is to let the relay decide on the hypothesis, and let the receiver simply follow this decision. It thus suffices that the relay forwards its decision to the receiver.  No quantization information is needed at the receiver.

\begin{corollary}\label{prop1c} Assume \eqref{eq:set1}--\eqref{eq:set3} hold.
	The exponent-rate region $\mathcal{E}^*(R,T)$ is the set of all nonnegative pairs $(\theta_y, \theta_z)$  that satisfy
	\begin{align}
	\theta_y &\leq I(U;\YH|\YC)\\
	\theta_z &\leq I(U;\YH|\YC),
	\end{align}
	for an auxiliary random variable $U$ satisfying the Markov chain  $U\to X \to (Y,Z)$  and the rate constraint
	\begin{align}
	R &\geq I(U;X|\YC),
	\end{align}
	where $Y=(\YC,\YH)$ and $(X,\YC,\YH,\ZC, \ZH) \sim  P_{X\YC\YH Z}$.
	(No constraint involves the rate $T$.)
\end{corollary}
\begin{IEEEproof} Achievability follows by specializing Proposition~\ref{cor-bin-state} to  $V$ being a constant and simplifying the expressions. The converse is similar to the proof of the converse to Corollary~\ref{prop1b}. \end{IEEEproof}

\begin{remark}
If we set $\YC$ to a constant, then the problem becomes one where
\begin{IEEEeqnarray}{rCl}
P_{XYZ} = P_{X|Y} \cdot P_{YZ}\\
Q_{XYZ} = P_X \cdot P_{YZ}.
\end{IEEEeqnarray}
The exponent-rate region then becomes the set of all nonnegative pairs $(\theta_y, \theta_z)$  that satisfy
	\begin{align}
	\theta_y &\leq I(U;Y)\\
	\theta_z &\leq I(U;Y),
	\end{align}
	for an auxiliary random variable $U$ satisfying the Markov chain  $U\to X \to (Y,Z)$  and the rate constraint
	\begin{align}
	R &\geq I(U;X).
	\end{align}
	The region is achievable using Proposition~\ref{prop:Markov}.
\end{remark}

\vspace{3mm}

\section{The Special Case ``$X^n\to Z^n \to Y^n$ Under Both Hypotheses"}\label{sec:MarkovXZY}

We consider a setup where $X^n \to Z^n \to Y^n$ forms a Markov chain under both hypotheses. {This setting models a situation where  the receiver lies in between the transmitter and the relay, and thus the signals at the sensor and the relay are conditionally independent given the signal at the receiver (decision center).} The two-hop network can still be an adequate communication model if all the communication from the transmitter to the receiver needs to be directed {through} the relay. This is for example the case in cellular systems where the relay is associated with a base station.

We treat two special cases: 1) same $P_{YZ}$ under both hypotheses, and 2) same $P_{XZ}$ under both hypotheses. Combined with the Markov chain $X \to Z \to Y$, these assumptions seem to suggest that the receiver cannot improve its error exponent by learning information about the observations  at  the relay (for case 1) or about the observations at the transmitter (for case 2). As we shall see, this holds only if the rates of communication are sufficiently high.  Otherwise, information about observations at both the transmitter and the relay can be combined to reduce the required rate of communication and thus also improve the performance of the system.
In this section we shall not employ binning, i.e., all achievability results below follow from Theorem~\ref{cascadeachnb}.

\bigskip
\subsection{Special Case 1: Same $P_{YZ}$ under both Hypotheses}


Consider first the setup where the pmfs $P_{XYZ}$ and $Q_{XYZ}$ decompose as
\begin{align}
P_{XYZ} &= P_{X|Z}\cdot P_{YZ},\label{case3a}\\
Q_{XYZ}&=P_X\cdot P_{YZ}.\label{case3b}
\end{align}
Since the pair of sequences $(Y^n,Z^n)$ has the same joint distribution under both hypotheses, no positive error exponent $\theta_z$ is possible when the message~$B$ sent from the relay to the receiver is only a function of $Y^n$ but not of the incoming message $M$. The structure of \eqref{case3a} and \eqref{case3b} might even suggest that $Y^n$ was not useful at the receiver and that the relay should simply forward a function of its incoming message $M$.  
Proposition~\ref{thm4} shows that this strategy is optimal when $T\geq R$, i.e., when the relay can forward the entire message to the receiver. On the other hand, Example~\ref{example1} shows that it can be suboptimal when $T<R$. 
\vspace{3mm}
\begin{proposition}\label{thm4} Assume conditions (\ref{case3a}) and (\ref{case3b}) and 
	\begin{equation}
T\geq R.\label{eq:TlargerR}
	\end{equation} 
	Then the exponent-rate region $\mathcal{E}(R,T)$ is  the set of all nonnegative pairs $( \theta_y,\theta_z)$ that satisfy
	\begin{align}
	\theta_y &\leq I(S;Y)\\
	\theta_z &\leq I(S;Z)\label{thetarec}
	\end{align}
	for some auxiliary random variable $S$ satisfying the Markov chain $S\to X\to (Y,Z)$ and the rate constraint
	\begin{align}
	 R\geq I(S;X),
	\end{align}
	where  $(X,Y,Z) \sim  P_{X|Z}\cdot P_{YZ}$.
\end{proposition}
\begin{IEEEproof} For achievability, specialize Theorem \ref{cascadeachnb} to $S=U=V$.  
	  The converse is proved in Appendix~\ref{convthm4}.
\end{IEEEproof}
\vspace{3mm}

We next consider an example that satisfies assumptions~\eqref{case3a} and \eqref{case3b}, but not \eqref{eq:TlargerR}. We assume $R\ge H(X),$ so the transmitter can reliably describe the sequence $X^n$ to the relay. 
When $T \geq R$, by Proposition~\ref{thm4}, the optimal strategy at the relay is to forward the incoming message $B=M$, i.e., to describe the entire $X^n$ to the receiver. 
In this example, to achieve the same exponent, it suffices that the relay describes only part of $X^n$, the choice of which depends on the relay's own observations $Y^n$. Thus, the relay only requires a rate $T$ that is smaller than $R$. 
\vspace{3mm}

\begin{example}\label{example1} Let under both hypotheses $\mathcal{H}=0$ and $\mathcal{H}=1$:
	\begin{equation}
	X\sim \mathcal{B}(1/2) \quad \textnormal{and} \quad Y \sim\mathcal{B}(1/2)\nonumber 
	\end{equation}
	be independent of each other.
	Also, let $N \sim \mathcal{B}(1/2)$  be independent  of the pair $(X,Y)$, and  
	$$ Z=(Z',Y) \qquad \textnormal{where} \quad Z'=\begin{cases} X& \textnormal{if }Y=0 \textnormal{ and } \mathcal{H}=0 \\  N & \textnormal{otherwise.} 
	\end{cases}. 
	$$
	Let $P_{XYZ}$ denote the joint pmf under $\mathcal{H}=0$ and $Q_{XYZ}$ the joint pmf under $\mathcal{H}=1$. 
	
	Notice that the triple  $(X,Y,Z)$ satisfies conditions (\ref{case3a}) and (\ref{case3b}). Moreover, since $P_{XY}=Q_{XY}$, the error exponent $\theta_y$ cannot be larger than  zero, and we   focus on the  error exponent $\theta_z$ achievable at the receiver. Notice that the conditional pmf 
	\begin{equation}\label{eq:same}
	P_{XZ|Y=1}= Q_{XZ|Y=1}.
	\end{equation}  The idea of our scheme is thus that the relay describes only the symbols 
	\begin{equation}\{X_t \colon t\in\{1,\ldots, n\},Y_t=0 \}
	\end{equation}
to the receiver. All other symbols are useless for distinguishing the two hypotheses.
Specifically, we specialize the scheme  in Subsection~\ref{sec:scheme} to the choice of random variables 
\begin{subequations}\label{eq:choice_auxiliary}
\begin{IEEEeqnarray}{rCl}
S &  &\textnormal{a constant}\\
U & =& X\\
 V& = & \begin{cases}U & \textnormal{if } Y=0 \\
U'& \textnormal{otherwise,}
\end{cases}
\end{IEEEeqnarray}
\end{subequations}
where $U'\sim \mathcal{B}(1/2)$ is independent of all other random variables.
Evaluating Theorem~\ref{cascadeachnb} for this choice
 proves achievability of the following error exponent at the receiver:  
 \begin{align}
 \label{eq:th}
& \hspace{-1.5cm}\min_{\substack{\tilde{P}_{VXYZ}:\\\tilde{P}_{VXY}=P_{VXY}\\\tilde{P}_{VZ}=P_{VZ}}} D(\tilde{P}_{VXYZ}\|P_{V|XY}Q_{XYZ}) \nonumber\\& \stackrel{(a)}{\geq } D({P}_{VZ}\|Q_{VZ}) \\
  &=D({P}_{VYZ'}\| Q_{VYZ'}) \\
    &\stackrel{(b)}{= } \mathbb{E}_{P_Y}[ D({P}_{VZ'|Y}\| Q_{VZ'|Y})] \\
  &\stackrel{(c)}{= } P_Y(0) \cdot D({P}_{VZ'|Y=0}\| Q_{VZ'|Y=0})  \\
 & = P_Y(0)\cdot  I(Z';V|Y=0)\\
  & = P_Y(0)\cdot  I(X;V|Y=0)\\
& = 1/2 H(X) =1/2, 
 \end{align}
 where the pmfs $P_{VXY}$, $P_{VZ}$, $P_{VYZ'}$ and the pmfs $Q_{VZ}$, $Q_{VYZ'}$ are obtained from the  definitions in \eqref{eq:choice_auxiliary} and the pmfs $P_{XYZ}$ and $Q_{XYZ}$, respectively, and mutual informations are calculated according to the joint pmf $P_{VXYZ'}$ defined through \eqref{eq:choice_auxiliary} and $P_{XYZ}$. 
 In the above, $(a)$ holds 
 	by the data-processing inequality, and by the second condition in the minimization; $(b)$ holds by the chain rule of KL-divergence and because $P_Y=Q_Y$; and $(c)$ holds because  $Q_{VZ'|Y=0}=P_{V|Y=0}\cdot P_{Z'|Y=0}$ whereas $Q_{VZ'|Y=1}=P_{VZ'|Y=1}$.
 
 The scheme requires rates 
 $$R=H(X)=1$$ and 
 \begin{IEEEeqnarray*}{rCl}\
 T & = & I(V;Y,U) \stackrel{(d)}{=}  I(V;X|Y) \\\
  &\stackrel{(e)}{=}& P_Y(0) \cdot I(V;X|Y=0) = 1/2,
  \end{IEEEeqnarray*} 
 where $(d)$ holds because $V$ is independent of $Y$ and $(e)$ holds because $V$ is also independent of $X$ unless $Y=0$. 
 
The error exponent in \eqref{eq:th} coincides with the largest exponent $D(P_{XYZ} \| Q_{XYZ})$ that is possible even in a fully centralized setup. We argue in the following that, provided $R=1$ and $T<1$, this error exponent cannot be achieved when the relay simply sends a function of the message $M$ to the receiver.
Notice that the setup incorporating only the transmitter and the receiver is a standard ``testing against independence" two-terminal setup \cite{Csiszar86} with  largest possible exponent equal to:
\begin{eqnarray}
&&\hspace{-2cm}\max_{P_{S|X} \colon T \geq I(S;X)} I(S;Z) \nonumber\\&\stackrel{(f)}{=} & \max_{P_{S|X} \colon T \geq I(S;X)}  I(S;Z'|Y)\nonumber\\
&=&1- \min_{P_{S|X} \colon T \geq I(S;X)}  H(Z'|Y,S)\nonumber\\
&=&1- \min_{P_{S|X} \colon H(X|S)\geq 1-T} \frac{1}{2}\cdot H(X|S)-\frac{1}{2}\nonumber\\
& \leq & \frac{1}{2} T,\label{th1}
\end{eqnarray}
where 
$(f)$ holds because $Z=(Z',Y)$ and because $(X,S)$ are independent of $Y$. This shows that the optimal exponent $1/2$ cannot be achieved if the relay simply sends a function of the incoming message whenever $T< 1$.

\end{example}
\bigskip
\subsection{Special Case 2: Same $P_{XZ}$ under both Hypotheses}
\vspace{3mm}
Consider next  a setup where
\begin{align}
P_{XYZ} &= P_{XZ}\cdot P_{Y|Z},\label{case2a}\\
Q_{XYZ} &= P_{XZ}\cdot P_Y.\label{case2b}
\end{align}
Notice that the pair of sequences $(X^n, Z^n)$ has the same joint pmf under both hypotheses. Thus, when the relay simply forwards the incoming message $M$ without conveying additional information about its observation $Y^n$ to the receiver, no positive error exponent $\theta_z$ is possible. On the contrary, as the following proposition shows, if
\begin{equation}\label{eq:const_T}
T \geq H(Y),
\end{equation}
then forwarding message $M$ to the receiver is useless, and  it suffices that the message $B$ sent from the relay to the receiver describes $Y^n$. In other words, under constraint \eqref{eq:const_T}, the optimal  error exponent $\theta_z$  coincides with the optimal error exponent of a point-to-point system that consists only of the relay and the receiver. The three-terminal multi-hop setup with a  transmitter observing $X^n$ can however achieve larger error exponent $\theta_z$ than the point-to-point system when \eqref{eq:const_T} does not hold. This is shown through  
 Example~\ref{ex:2} ahead.
\bigskip 

\begin{proposition}\label{thm3} Assume (\ref{case2a})--\eqref{eq:const_T}. Under these assumptions, the exponent-rate region $\mathcal{E}^*(R,T)$ is  the set of all nonnegative pairs $( \theta_y,\theta_z)$ that satisfy
	\begin{align}
	\theta_y &\leq I(U;Y),\label{thm3aa}\\
	\theta_z &\leq I(Y;Z),\label{thm3bb}
	\end{align}
	for some auxiliary random variable $U$ satisfying the Markov chain $U\to X\to (Y,Z)$ and the rate constraint
	\begin{align}
	R &\geq I(U;X),\label{thm3cc}	
	\end{align}
\end{proposition}
where  $(X,Y,Z) \sim  P_{XZ}\cdot P_{Y|Z}$.
\begin{IEEEproof} Achievability follows by specializing Theorem \ref{cascadeachnb} to $S=U$ and $V=Y$. The converse for \eqref{thm3aa} is the same as in the point-to-point setting (without receiver). The converse for \eqref{thm3bb} follows by Stein's lemma (without communication constraints) \cite{Cover}.
\end{IEEEproof}

\bigskip
We next consider an example where assumptions \eqref{case2a} and \eqref{case2b} hold, but not~\eqref{eq:const_T}. 
\bigskip

\begin{example}\label{ex:2}
 Let under both hypotheses $\mathcal{H}=0$ and $\mathcal{H}=1$:	\begin{equation}
	X\sim \mathcal{B}(1/2) \quad \textnormal{and} \quad Y \sim \mathcal{B}(1/2)\nonumber 
	\end{equation}
	be independent of each other.
	Also, let $N \sim \mathcal{B}(1/2)$  be independent of the pair $(X,Y)$, and  
	$$ Z=(Z',X) \qquad \textnormal{where} \quad Z'=\begin{cases} Y & \textnormal{if }X=0 \textnormal{ and } \mathcal{H}=0 \\  N & \textnormal{otherwise.} 
	\end{cases}
	$$
The described triple  $(X,Y,Z)$ satisfies conditions (\ref{case2a}) and (\ref{case2b}). 
Moreover, since the pmf of the sequences  $(X^n,Y^n)$ is the same under both hypotheses, the best error exponent $\theta_y$ is zero, so we focus on the receiver's error exponent $\theta_z$. 	By Proposition~\ref{thm3}, the largest error exponent $\theta_z$  that is achievable is \begin{equation}
\label{eq:optimal}
\theta_z^\star=I(Y;Z)=I(Y;Z'|X)=1/2.
\end{equation} 
As we show in the following, $\theta^\star_z$ is achievable with $T=1/2$. To see this, notice that 
\begin{equation}
	P_{YZ|X=1}= Q_{YZ|X=1}.
\end{equation}
It thus suffices that the relay conveys the values of its observations $\{Y_t \colon t\in\{1,\ldots,n\},X_t=0\}$ to the receiver. This is achieved by specializing the coding and testing scheme of Subsection~\ref{sec:scheme} to the choice of $S$ being a constant and 
\begin{IEEEeqnarray*}{rCl}
U & = & \begin{cases}0 & \textnormal{if } X=0 \\
1 & \textnormal{otherwise}
\end{cases} \\
 V& = & \begin{cases}Y& \textnormal{if } U=0 \\
Y' & \textnormal{otherwise},
\end{cases}
\end{IEEEeqnarray*}
where $Y'\sim\mathcal{B}(1/2)$ is independent of $(X,Y,Z)$.
By  Theorem \ref{cascadeachnb}, the scheme requires rates 
 $$R=I(U;X)=H(U)=1$$ and $$T=I(V;Y,U)= P_U(0)\cdot I(V;Y|U=0) = 1/2.$$
 It achieves the optimal error exponent $\theta_z^\star$ in \eqref{eq:optimal}:
  \begin{IEEEeqnarray}{rCl}
  \lefteqn{\min_{\substack{\tilde{P}_{UVXYZ}  \colon \\ \tilde{P}_{UX}=P_{UX} \\ \tilde{P}_{UVY}=P_{UVY} \\ \tilde{P}_{VZ} = P_{VZ}}} D( \tilde{P}_{UVXYZ} \| P_{U|X}  P_{V|UY} Q_{XYZ})} \qquad  \nonumber \\
   & \stackrel{(a)}{\geq} & D(P_{VZ} \| Q_{VZ}) \\
   & = &  \mathbb{E}_{{P}_{X}}\big[  D({P}_{VZ'|X} \|    Q_{VZ'|X}) \big] \nonumber \\
   & \stackrel{(b)}= & P_X(0) D({P}_{VZ'|X=0} \|    P_{V|X=0} P_{Z'|X=0})\nonumber \\
   &= & P_X(0) I(V;Z'|X=0) \nonumber \\
   &=&  1/2,\label{eq:theta}
 \end{IEEEeqnarray} 
 where $(a)$ holds by the data-processing inequality for KL-divergences and by defining $Q_{VZ}$ to be the marginal of the joint pmf $P_{U|X}P_{V|UY}Q_{XYZ}$; and $(b)$ holds because $Q_{VZ'|X=0}=P_{V|X=0}P_{Z'|X=0}$ whereas $Q_{VZ'|X=1}=P_{VZ'|X=1}$.

 Using similar arguments as in Example~\ref{example1}, it can be shown that the optimal error exponent $\theta^\star_z$ in \eqref{eq:optimal} cannot be achieved without the transmitter's help when $T<1$. 
 \end{example}

\section{Concluding Remarks}\label{sec:conclusion}

The paper presents  coding and testing schemes for a two-hop relay network, and  the corresponding exponent-rate region. The schemes combine cascade source coding with a unanimous decision-forwarding strategy where the receiver decides on the null hypothesis only if both the transmitter and relay have decided on it. The schemes are shown to attain the entire exponent-rate region for some cases of testing against independence or testing against conditional independence when the Markov chain $X^n\to Y^n\to Z^n$ holds. In these cases, the source coding part of our  coding schemes simplifies to independent source codes for the transmitter-to-relay link and for  the relay-to-receiver link. The proposed schemes are also shown to be optimal in some special cases when the Markov chain $X^n\to Z^n\to Y^n$ holds. For large enough communication rates and when testing against independence, it is again optimal to employ independent source codes for the two links. But, when the rate on the relay-to-receiver link is small, this simplification can be suboptimal.  

One of our coding schemes employs binning to decrease the required rates of communication. Binning makes the characterization of the achievable exponent region much more involved. For the proposed scheme we have two competing exponents for the error exponent at the relay and four competing exponents for the error exponent at the receiver. Notice that, in our scheme, we only bin the satellite codebooks but not the cloud-center codebooks. Further performance improvement might be obtained by binning also the cloud center; this would however lead to an expression with ten competing exponents at the receiver. 
                                                                                                                                              
                                                                                                                                           

\section{Acknowledgement}

The authors would like to thank Pierre Escamilla and Abdellatif Zaidi for  helpful discussions, and Associate Editor Shun Watanabe and the anonymous reviewers for their valuable comments.

\bibliographystyle{IEEEtran}
\bibliography{references}

\appendices

\section{Proof of Theorem \ref{cascadeachnb}}\label{cascadeachnbpr}
We bound the probabilities of error of the scheme averaged over the  random code construction $\mathcal{C}$.
The analysis of the error probabilities at the relay is standard. We therefore focus on the error probabilities at the receiver.

If $M\neq 0$ and $B\neq 0$, let $I,J,K$ be the random indices sent over the bit pipes and define the following  events:
\begin{align*}
\mathcal{E}_{\textnormal{Relay}}:& \quad\big\{(S^n(I),U^n(J|I), Y^n ) \notin\mathcal{T}_{\mu/2}^n(P_{SUY})\big\},\\
\mathcal{E}_{\textnormal{Rx}}:& \quad\big\{(S^n(I),U^n(J|I), V^n(K|I), Z^n ) \notin\mathcal{T}_{\mu}^n(P_{SUVZ})\big\}.
\end{align*}

The type-I error probability at the receiver averaged over the random code construction can be bounded, for large enough $n$, as follows
\begin{IEEEeqnarray}{rCl}
\mathbb{E}_{\mathcal{C}}\big[	\alpha_{z,n} \big]& \leq & \Pr[ M=0 \; \textnormal{or}\; B=0\; \textnormal{or}\;   \mathcal{E}_{\textnormal{Relay}} \; \textnormal{or} \;   \mathcal{E}_{\textnormal{Rx}} | \mathcal{H}=0] \nonumber \\
	& \leq & \Pr[M=0| \mathcal{H}=0] \nonumber\\*&&{}+ \Pr[B=0\; \textnormal{or}\;   \mathcal{E}_{\textnormal{Relay}}|M\neq 0, \mathcal{H}=0] \nonumber\\*&&{}+ \Pr[\mathcal{E}_{\textnormal{Rx}} |M\neq 0, B\neq 0, \mathcal{H}=0]   \\
	& \stackrel{(a)}{\leq} & \epsilon/32 +   \Pr[\mathcal{E}_{\textnormal{Relay}}|M\neq 0, \mathcal{H}=0]  \nonumber\\&&{}+\Pr[B=0| M \neq 0,\mathcal{E}_{\textnormal{Relay}}^c, \mathcal{H}=0]   +   \epsilon/32  \nonumber\\\\
	& \stackrel{(b)}{\leq} & \epsilon/32 + \epsilon/32 + \epsilon/32 +\epsilon/32 \\
	& = &\epsilon/8,
\end{IEEEeqnarray}
where $(a)$ holds by the covering lemma and the rate constraint \eqref{casnb1}, and both $(a)$ and $(b)$ hold  by the Markov lemma \cite{ElGamal}. 

We now bound the probability of type-II error at the receiver. 
Let $\mathcal{P}^n$ be the set of all types over the product alphabets $ \mathcal{S}^n \times \mathcal{U}^n\times \mathcal{V}^n \times\mathcal{X}^n \times \mathcal{Y}^n \times \mathcal{Z}^n$. Also, let 
$\mathcal{P}^n_\mu$ be the subset of types $\pi_{SUVXYZ} \in \mathcal{P}^n$ that simultaneously satisfy the following three conditions:
\begin{align}|\pi_{SUX}-P_{SUX}| & \leq \mu/4,\\|\pi_{SUVY}-P_{SUVY}| & \leq \mu/2,\\
|\pi_{SVZ}-P_{SVZ}| & \leq \mu.
\end{align}

Now, consider the type-II error probability averaged over the random code construction. For all
$(i,j,k)\in\{1,\ldots ,\lfloor 2^{nR_s} \rfloor \}\times \{1,\ldots,\lfloor 2^{nR_u}\rfloor\}\times \{1,\ldots,\lfloor 2^{nR_v}\rfloor\}$, define the events:
\begin{IEEEeqnarray}{rCl}
	\mathcal{E}_{\text{Tx}}(i,j)&=&\left\{ (S^n(i),U^n(j|i),X^n)\in\mathcal{T}_{\mu/4}^n(P_{SUX}) \right\},\nonumber\\\\
	\mathcal{E}_{\text{Rel}}(i,j,k)&=&\nonumber\\&&\hspace{-0.9cm}\left\{ (S^n(i),U^n(j|i),V^n(k|i),Y^n)\in\mathcal{T}_{\mu/2}^n(P_{SUVY}) \right\},\nonumber\\\\
	\mathcal{E}_{\text{Rx}}(i,k)&=&\left\{ (S^n(i),V^n(k|i),Z^n)\in\mathcal{T}_{\mu}^n(P_{SVZ}) \right\}.
	\end{IEEEeqnarray}
We have
\begin{align}
\mathbb{E}_{\mathcal{C}}\big[\beta_{z,n}\big] &=\Pr\left[ \hat{\mathcal{H}}_z=0 \Big| \mathcal{H}=1 \right]\nonumber\\&\hspace{-1.2cm}\leq \Pr \Big[\bigcup_{i,j,k}\; \left(\mathcal{E}_{\text{Rx}}(i,k) \;\cap\;\mathcal{E}_{\text{Rel}}(i,j,k)\; \cap \mathcal{E}_{\text{Tx}}(i,j)\right) \;\Big| \nonumber\\[-3mm]
&  \hspace{4cm}  \mathcal{H}=1\Big]. \label{eq:bound1}
\end{align}
(The inequality in \eqref{eq:bound1} comes from the fact that the transmitter chooses the pair of indices $(i,j)$ uniformly at random over all pairs for which event $\mathcal{E}_{\textnormal{Tx}}(i,j)$ holds. There can thus exist a triple $(i',j',k')$ satisfying  $\left(\mathcal{E}_{\text{Rx}}(i',k') \;\cap\;\mathcal{E}_{\text{Rel}}(i',j',k')\; \cap \mathcal{E}_{\text{Tx}}(i',j')\right)$ but the receiver still decides on $\hat{\mathcal{H}}_z=1$  because the transmitter chose a pair $(i,j)$ for which  $(\mathcal{E}_{\text{Rel}}(i,j,k)\; \cap \mathcal{E}_{\text{Tx}}(i,j))$ is violated for all values of $k$.)

We continue by applying the union bound: 
\begin{IEEEeqnarray}{rCl}
	&&\hspace{0cm}\Pr\Big[  \;\bigcup_{i,j,k}\; \left(\mathcal{E}_{\text{Rx}}(i,k) \;\cap\;\mathcal{E}_{\text{Rel}}(i,j,k)\; \cap \mathcal{E}_{\text{Tx}}(i,j)\right) \;\Big|\; \mathcal{H}=1 \Big]\nonumber\\[2ex]
	&&\leq \sum_{i,j,k} \Pr\left[   \mathcal{E}_{\text{Rx}}(i,k) \;\cap\;\mathcal{E}_{\text{Rel}}(i,j,k)\; \cap \mathcal{E}_{\text{Tx}}(i,j) \;\Big|\; \mathcal{H}=1 \right]\nonumber\\
	&&= \sum_{i,j,k} \Pr\Bigg[   (S^n(i),V^n(k|i),Z^n)\in\mathcal{T}_{\mu}^n(P_{SVZ}), \nonumber\\&&\hspace{1.9cm} (S^n(i),U^n(j|i),V^n(k|i),Y^n)\in\mathcal{T}_{\mu/2}^n(P_{SUVY}),\nonumber\\&&\hspace{2.1cm} (S^n(i),U^n(j|i),X^n)\in\mathcal{T}_{\mu/4}^n(P_{SUX})\Big|\; \mathcal{H}=1 \Bigg]\nonumber\\
	&& = \sum_{i,j,k}\;\sum_{\substack{\pi_{SUVXYZ} \\\in \mathcal{P}^n_{\mu}}}\hspace{-0.2cm}\Pr\Bigg[\text{tp}\left( S^n(i), U^n(j|i), V^n(k|i),X^n, Y^n, Z^n\right)\nonumber\\[-2ex]&&\hspace{5.6cm} = \pi_{SUVXYZ}\Big| \mathcal{H}=1\Bigg]\nonumber\\[1.5ex]
	&&\leq 2^{n(R_s+R_u+R_v)}\cdot   \;\;(n+1)^{|\mathcal{S}|\cdot|\mathcal{U}|\cdot|\mathcal{V}|\cdot|\mathcal{X}|\cdot|\mathcal{Y}|\cdot|\mathcal{Z}|}\nonumber\\&&\hspace{0.5cm}\cdot \max_{\pi_{USVXYZ}\in\mathcal{P}_\mu^n} 2^{-n(D(\pi_{SUVXYZ} \| P_{SU}P_{V|S}Q_{XYZ})-\mu)},
	\end{IEEEeqnarray}
where the last inequality holds by Sanov's theorem \cite{Cover}. Indeed, by the code construction, the three codewords  $(S^n(i),U^n(j|i),$ $V^n(k|i))$ are drawn i.i.d. according to $P_{SU} P_{V|S}$. Furthermore, they are independent of $(X^n,Y^n, Z^n)$, which, under $\mathcal{H}=1$, are drawn i.i.d. according to $Q_{XYZ}$. Therefore, 
\begin{align}
\mathbb{E}_{\mathcal{C}}\big[\beta_{z,n}\big] \leq & (n+1)^{|\mathcal{S}|\cdot|\mathcal{U}|\cdot|\mathcal{V}|\cdot|\mathcal{X}|\cdot|\mathcal{Y}|\cdot|\mathcal{Z}|} \times\max_{\pi_{USVXYZ}\in\mathcal{P}_\mu^n} \nonumber\\&\hspace{0cm}   \Big[ 2^{n(R_s+R_u+R_v-D(\pi_{SUVXYZ} \| P_{SU}P_{V|S}Q_{XYZ})+\mu)}  \Big].\label{eq:lastbeta}
\end{align}
Plugging  the rate expressions \eqref{casnb1}--\eqref{casnb3} into \eqref{eq:lastbeta}  results in the following upper bound:
\begin{align}
\mathbb{E}_{\mathcal{C}}\big[\beta_{z,n}\big]  &\leq (n+1)^{|\mathcal{S}|\cdot|\mathcal{U}|\cdot|\mathcal{V}|\cdot|\mathcal{X}|\cdot|\mathcal{Y}|\cdot|\mathcal{Z}|}\cdot 2^{-n \theta_{z,\mu}},
\end{align}
where 
\begin{align}
\theta_{z,\mu} :{=} & \min_{\pi_{USVXYZ}\in\mathcal{P}^n_\mu} \big[ D(\pi_{SUVXYZ} \| P_{SU}P_{V|S}Q_{XYZ}) \nonumber\\&\hspace{2cm}-I(X;S,U)-I(Y,U;V|S)-\mu\big] .
\end{align}
Now, by simplifying terms and employing the continuity of KL-divergence, we conclude that 
\begin{equation}
\mathbb{E}_{\mathcal{C}}[ \beta_{z,n}]  \leq  2^{ - n (\theta_z - \delta_n(\mu))},
\end{equation}
for some function $\delta_n(\mu)$ that tends to $0$ as $n \to \infty$ and $\mu \to 0$, and
\begin{IEEEeqnarray}{rCl}
	\theta_z &:=&  \min_{\substack{\tilde{P}_{SUVXYZ}:\\\tilde{P}_{SUX}=P_{SUX}\\\tilde{P}_{SVUY}=P_{SVUY}\\\tilde{P}_{SVZ}=P_{SVZ}}} D(\tilde{P}_{SUVXYZ}\|P_{SU}P_{V|S}Q_{XYZ})\nonumber\\[-2ex]&&\hspace{2.5cm}{}-I(X;S,U)-I(Y,U;V|S)\nonumber\\[1ex]
	&=&\min_{\substack{\tilde{P}_{SUVXYZ}:\\\tilde{P}_{SUX}=P_{SUX}\\\tilde{P}_{SVUY}=P_{SVUY}\\\tilde{P}_{SVZ}=P_{SVZ}}}\;\; \sum_{s,u,v,x,y,z}\Bigg[\tilde{P}_{SUVXYZ}(s,u,v,x,y,z)\times \nonumber\\*[-2ex]&&\hspace{2cm}\log \frac{\tilde{P}_{SUVXYZ}(s,u,v,x,y,z)}{P_{SU}(s,u)P_{V|S}(v|s)Q_{XYZ}(x,y,z)}\nonumber\\*[1ex] &&\hspace{1.5cm}{}-P_{SUX}(s,u,x)\log \frac{P_{SU|X}(s,u|x)}{P_{SU}(s,u)}\nonumber\\* &&\hspace{1.5cm}{}-P_{SUVY}(s,u,v,y)\log \frac{P_{V|SUY}(v|s,u,y)}{P_{V|S}(v|s)}\Bigg]\nonumber\\
			&\stackrel{(c)}{=}&\min_{\substack{\tilde{P}_{SUVXYZ}:\\\tilde{P}_{SUX}=P_{SUX}\\\tilde{P}_{SVUY}=P_{SVUY}\\\tilde{P}_{SVZ}=P_{SVZ}}}\;\; \sum_{s,u,v,x,y,z} \tilde{P}_{SUVXYZ}(s,u,v,x,y,z)\times\nonumber\\&&\log \frac{\tilde{P}_{SUVXYZ}(s,u,v,x,y,z)}{P_{SU|X}(s,u|x)P_{V|SUY}(v|s,u,y)Q_{XYZ}(x,y,z)}\nonumber\\
			&=&\min_{\substack{\tilde{P}_{SUVXYZ}:\\\tilde{P}_{SUX}=P_{SUX}\\\tilde{P}_{SVUY}=P_{SVUY}\\\tilde{P}_{SVZ}=P_{SVZ}}} D\left(\tilde{P}_{SUVXYZ}\| P_{SU|X}P_{V|SUY}Q_{XYZ} \right),\nonumber\\
	\end{IEEEeqnarray}
where $(c)$ follows from the first and second constraints on the minimization and by re-arranging terms.

To summarize, we showed that 
on average (over the random codebook constructions $\mathcal{C}$) and for sufficiently large $n$:
\begin{IEEEeqnarray}{rCl}
\mathbb{E}_{\mathcal{C}}[ \alpha_{z,n}]  &\leq &\frac{\epsilon}{8} \\
\mathbb{E}_{\mathcal{C}}[ \beta_{z,n}]  &\leq&  2^{ - n (\theta_z - \delta_n(\mu))}.
\end{IEEEeqnarray}  
Similar arguments can be employed to show that also 
\begin{IEEEeqnarray}{rCl}
\mathbb{E}_{\mathcal{C}}[ \alpha_{y,n}]  &\leq &\frac{\epsilon}{4} \\
\mathbb{E}_{\mathcal{C}}[ \beta_{y,n}]  &\leq&  2^{ - n (\theta_y - \tilde{\delta}_n(\mu))},
\end{IEEEeqnarray}   for some function $\tilde{\delta}_n(\mu)$ that tends to 0 as $n \to \infty$ and as $\mu \to 0$, and for 
\begin{equation}
\theta_y :=
\min_{\substack{\tilde{P}_{SUXY}:\\\tilde{P}_{SUX}=P_{SUX}\\\tilde{P}_{SUY}=P_{SUY}}} D\left(\tilde{P}_{SUXY}\| P_{SU|X}Q_{XY} \right).
\end{equation}

We now argue that for all sufficiently large blocklengths  $n$ there must exist at least one deterministic code construction $\mathcal{C}^*_n$ and a function  $\hat{\delta}_n(\mu)$ that tends to 0 as $n \to \infty$ and as $\mu \to 0$, such that for this code:
\begin{subequations}\label{eq:requirements}
\begin{IEEEeqnarray}{rCl}
 \alpha_{y,n}  &\leq &\epsilon \\
\alpha_{z,n} &\leq & \epsilon \\
\beta_{y,n}  &\leq&  2^{ - n (\theta_y - \hat{\delta}_n(\mu))}\\
 \beta_{z,n}  &\leq&  2^{ - n (\theta_z - \hat{\delta}_n(\mu))}.
\end{IEEEeqnarray}  
\end{subequations}
To this end, we start by eliminating a set of code constructions that yield largest $\alpha_{y,n}$. The size of the set is chosen such that its total probability is at least  $1/2$ and at most  $3/4$. (Instead of $3/4$, one can choose a value that is arbitrarily close to $1/2$. Such a choice is always feasible for sufficiently large blocklengths $n$, because the maximum probability of a single code construction tends to 0 as $n \to \infty$  unless all random variables are constants, but this latter case is not interesting.) 
Each of the code constructions in the remaining set $\mathcal{C}_1$ then has probability of type-I error
\begin{equation}
\alpha_{y,n}\leq  \frac{\epsilon}{4} \cdot \frac{4}{3} = \frac{\epsilon}{3}
\end{equation}
and on average these code constructions have probability of  type-I error and type-II errors
\begin{IEEEeqnarray}{rCl}
\mathbb{E}_{\mathcal{C}_1}[ \alpha_{z,n}]  &\leq &\frac{\epsilon}{8}\cdot\frac{1}{1-\frac{3}{4}} = \frac{\epsilon}{2}  \\
\mathbb{E}_{\mathcal{C}_1}[ \beta_{z,n}]  &\leq&  2^{ - n (\theta_z - \delta_n(\mu))} \cdot \frac{1}{1-\frac{3}{4}} \\
\mathbb{E}_{\mathcal{C}_1}[ \beta_{y,n}]  &\leq&  2^{ - n (\theta_y - \delta_n(\mu))} \cdot \frac{1}{1-\frac{3}{4}}.
\end{IEEEeqnarray}
In the same way we continue to eliminate a subset of $\mathcal{C}_1$ containing the code constructions with largest $\alpha_{z,n}$ such that the probability of this subset is at least  $1/2$ and at most  $3/4$ the probability of $\mathcal{C}_1$. Call the remaining set $\mathcal{C}_2$. From $\mathcal{C}_2$, we then  eliminate code constructions that yield largest $\beta_{y,n}$, such that all the eliminated code constructions (in this step) constitute at least  $1/2$ and at most  $3/4$ the probability of $\mathcal{C}_2$. Finally, from the code constructions that survive all eliminations, we pick the one with the smallest $\beta_{z,n}$. 
This finally selected code $\mathcal{C}^*$ then satisfies
\begin{IEEEeqnarray}{rCl}
 \alpha_{y,n}  &\leq &\frac{\epsilon}{3} \\
\alpha_{z,n} &\leq &  \frac{\epsilon}{2} \cdot \frac{4}{3} = \frac{2}{3} \epsilon \\
\beta_{y,n}  &\leq&  2^{ - n (\theta_y -{\delta}_n(\mu))}  \cdot \left(\frac{1}{1- \frac{3}{4}}\right)^2 \cdot \frac{4}{3} =2^{ - n (\theta_y -{\delta}_n(\mu))} \cdot \frac{64}{ 3} \nonumber\\\\
 \beta_{z,n}  &\leq&  2^{ - n (\theta_z - \tilde{\delta}_n(\mu))}  \cdot \left(\frac{1}{1- \frac{3}{4}}\right)^3=2^{ - n (\theta_y -\tilde{\delta}_n(\mu))} \cdot 64.\nonumber\\
\end{IEEEeqnarray}  
If we set $\hat{\delta}_{n}(\mu) = \max\{\delta_n(\mu), \tilde{\delta}_{n}(\mu)\} + \frac{6}{n}$, then all inequalities \eqref{eq:requirements} are satisfied.

\section{Proof of Theorem \ref{thm_binning}}\label{thm1pr}
It only remains to prove \eqref{ach}.
We analyze the probabilities of error of the coding and testing scheme described in Subsection~\ref{sec:binning_scheme} averaged over the random code construction. By standard arguments (successively eliminating the worst half of the codebooks as described at the end of Appendix~\ref{cascadeachnbpr})  the desired result can be proved for a set of deterministic codebooks.

Fix an arbitrary $\epsilon>0$ and the scheme's parameter $\mu>0$. 
For a fixed blocklength $n$, 
	let $\mathcal{P}_{\mu,\text{type-I}}^n$ 
be the subset of types over the product alphabet $\mathcal{S}^n\times \mathcal{S}^n\times\mathcal{Y}^n$ that  satisfy the following  conditions for all $(s,s',y)\in \mathcal{S}\times\mathcal{S} \times \mathcal{Y}$:
\begin{align}
|\pi_{SY}(s,y)-P_{SY}(s,y)| & \leq \mu,\label{cona}\\
|\pi_{S'}(s')-P_S(s)|  &\leq \mu,\label{conb}\\
H_{\pi_{S'Y}}(S'|Y)&\leq H_{\pi_{SY}}(S|Y).\label{conc} 
\end{align}


Notice that, when we let $n\to\infty$ and then $\mu\to 0$, each element in $\mathcal{P}_{\mu,\text{type-I}}^n$ will approach an element of 
\begin{align}
&\mathcal{P}_{\text{type-I}}^* := \big\{ \tilde{P}_{SS'Y} \colon \tilde{P}_{SY}=P_{SY} \textnormal{ and } \tilde{P}_{S'}=P_S \textnormal{ and }\nonumber\\&\hspace{2.8cm} H_{\tilde{P}_{S'Y}}(S'|Y) \leq  H_{\tilde{P}_{SY}}(S|Y) \big\}.\label{eq:Pstar}
\end{align}
  We first analyze the type-I error probability {$\alpha_{y,n}$}. For the case of $M\neq 0$, let $L$  be the index chosen at the transmitter. 
  Define   events 
 \begin{align}
 \mathcal{E}_{\text{Tx}}^{(0)} &:= \big\{ ( S^n(m,\ell), X^n) \notin \mathcal{T}_{\mu/2}^n(P_{SX}),\ \forall (m,\ell) \big\},\\
 \mathcal{E}_{\text{Rx}}^{(1)} &:= \big\{  (S^n(M,L),Y^n)\notin \mathcal{T}_{\mu}^n(P_{SY}) \big\},\\
 \mathcal{E}_{\text{Rx}}^{(2)} &:= \big\{ \exists {\ell}'\neq L \colon \;\; S^n(M,\ell')\in \mathcal{T}_{\mu}^n(P_{S})\;\;\textnormal{and} \nonumber\\&\hspace{1cm}  H_{\text{tp}(S ^n(M,L),Y^n)}(S|Y)\geq  H_{\text{tp}(S^n(M,{\ell}'),Y^n)}(S|Y)   \big\}.
 \end{align}
For all sufficiently large $n$, 
 the  average type-I error probability can  be bounded as:
 \begin{align}
{\mathbb{E}_{\mathcal{C}}[ \alpha_{y,n} ]} &= \Pr\big[ \hat{\mathcal{H}}_{y}=1 \big| \mathcal{H}=0\big]\\
 & \leq \Pr\Big[\mathcal{E}_{\text{Tx}}^{(0)}\;\cup\;\mathcal{E}_{\text{Rx}}^{(1)}\;\cup\;\mathcal{E}_{\text{Rx}}^{(2)}\Big | \mathcal{H}=0\Big]\\
 &\leq \Pr\Big[\mathcal{E}_{\text{Tx}}^{(0)} \Big|  \mathcal{H}=0\Big] +  \Pr\Big[\mathcal{E}_{\text{Rx}}^{(1)}\Big|\mathcal{E}_{\text{Tx}}^{(0)c}, \mathcal{H}=0\Big]\nonumber\\&\hspace{0.5cm}+\Pr\Big[\mathcal{E}_{\text{Rx}}^{(2)}\Big|\mathcal{E}_{\text{Rx}}^{(1)c},\mathcal{E}_{\text{Tx}}^{(0)c}, \mathcal{H}=0\Big]\\
 &\stackrel{(a)}{\leq} \epsilon/6 + \Pr\Big[\mathcal{E}_{\text{Rx}}^{(1)}\Big|\mathcal{E}_{\text{Tx}}^{(0)c}, \mathcal{H}=0\Big]\nonumber\\&\hspace{0.5cm}+\Pr\Big[\mathcal{E}_{\text{Rx}}^{(2)}\Big|\mathcal{E}_{\text{Rx}}^{(1)c},\mathcal{E}_{\text{Tx}}^{(0)c}, \mathcal{H}=0\Big]\\
 &\stackrel{(b)}{\leq} \epsilon/6 + \epsilon/6 + \Pr\Big[\mathcal{E}_{\text{Rx}}^{(2)}\Big|\mathcal{E}_{\text{Rx}}^{(1)c},\mathcal{E}_{\text{Tx}}^{(0)c}, \mathcal{H}=0\Big]\\
 &  \stackrel{(c)}{\leq} \epsilon/6 + \epsilon/6 + \epsilon/6  \\
 & =\epsilon/2,
 \end{align}
where  inequality $(a)$ follows from the code construction; $(b)$ follows from the Markov lemma \cite{ElGamal};  and  $(c)$ is justified in what follows. 
Notice first that  by the symmetry of the codebook construction, when bounding the probability $\Pr\left[\mathcal{E}_{\text{Rx}}^{(2)}\Big|\mathcal{E}_{\text{Rx}}^{(1)c},\mathcal{E}_{\text{Tx}}^{(0)c}, \mathcal{H}=0\right]$, we can specify $M=L=1$ 
and proceed as: 
\begin{IEEEeqnarray}{rCl}
\lefteqn{ \Pr\left[\mathcal{E}_{\text{Rx}}^{(2)}\Big|\mathcal{E}_{\text{Rx}}^{(1)c},\mathcal{E}_{\text{Tx}}^{(0)c}, M=L=1, \mathcal{H}=0\right] }   \\[1ex]
	& \le & \sum_{\ell'=2}^{\lfloor 2^{nR'} \rfloor} \Pr\Big[
	S^n(1,\ell')\in\mathcal{T}_{\mu}^n(P_S)\;,\nonumber\\&&\hspace{1.3cm}
	 H_{\text{tp}(S^n(1,1),Y^n)}(S|Y) \geq   H_{\text{tp}(S^n(1,\ell'),Y^n)}(S|Y)\;  \Big| \nonumber\\
	&&\hspace{1.3cm} (S^n(1,1),Y^n)\in \mathcal{T}_{\mu}^n(P_{SY}),\; \nonumber\\&&\hspace{1.3cm} (S^n(1,1),X^n)\in \mathcal{T}_{\mu/2}^n(P_{SX}),\;\nonumber\\&&\hspace{1.3cm}  M=L=1,\;\; \mathcal{H}=0 \Big]\\[1.2ex]
 	& \le & \sum_{\ell'=2}^{\lfloor 2^{nR'} \rfloor} \Pr\Big[ H_{\text{tp}(S^n(1,1),Y^n)}(S|Y) \geq  H_{\text{tp}(S^n(1,{\ell'}),Y^n)}(S|Y) \big| \nonumber\\*
 	&&\qquad\qquad\;\; \;  (S^n(1,1),Y^n)\in \mathcal{T}_{\mu}^n(P_{SY}),\;\;\nonumber\\*&&\qquad\qquad\;\; \;  (S^n(1,1),X^n)\in \mathcal{T}_{\mu/2}^n(P_{SX}),\;\; \nonumber\\*&&\qquad\qquad\;\; \; S^n(1,{\ell'}) \in\mathcal{T}_{\mu}^n(P_S)),\;M=L=1,\; \mathcal{H}=0 \Big] \nonumber\\  \\[1.2ex]
 	&= & \sum_{\substack{\pi_{SS'Y}\\\in\mathcal{P}_{\mu,\text{type-I}}^{n}}}\; \; \sum_{{\ell'}=2}^{\lfloor 2^{nR'}\rfloor} \; \; \sum_{\substack{s^n,s'^n,y^n:\\\text{tp}(s^n,s'^{n},y^n)\\=\pi_{SS'Y}}} \nonumber\\*&&\hspace{1cm}\Pr\Big[S^n(1,1)=s^n, S^n(1,{\ell'})=s'^n,Y^n=y^n\; \big| \;\nonumber \\*		
 	&& \qquad\qquad\;\;\; (S^n(1,1),Y^n)\in \mathcal{T}_{\mu}^n(P_{SY}),\nonumber \\		
 	&& \qquad\qquad\;\;\; (S^n(1,1),X^n)\in \mathcal{T}_{\mu/2}^n(P_{SX}),\; \;\nonumber \\		
 	&& \qquad\qquad\;\;\;  S^n(1,{\ell'}) \in\mathcal{T}_{\mu}^n(P_S) ),\;\;M=L=1,\;\; \mathcal{H}=0\Big] \nonumber \\
 	 \\[1.2ex]
 	&\stackrel{(d)}{=}& \sum_{\substack{\pi_{SS'Y}\\\in\mathcal{P}_{\mu,\text{type-I}}^{n}}} \;\; \sum_{{\ell'}=2}^{\lfloor 2^{nR'}\rfloor}\;\sum_{\substack{s^n,s'^n,y^n:\\\text{tp}(s^n,s'^{n},y^n)\\=\pi_{SS'Y}}} \nonumber\\&&\qquad\;\;\;\Pr\Big[S^n(1,1)=s^n,Y^n=y^n\; \big| \nonumber\\
 	&&\qquad\qquad\;\;\;   (S^n(1,1),Y^n)\in \mathcal{T}_{\mu}^n(P_{SY}),\;\; \nonumber\\
 	&&\qquad\qquad\;\;\;  (S^n(1,1),X^n)\in \mathcal{T}_{\mu/2}^n(P_{SX}),\; \;\nonumber\\
 	&&\qquad\qquad\;\;\;  S^n(1,{\ell'}) \in\mathcal{T}_{\mu}^n(P_S)),\;M=L=1,\; \mathcal{H}=0 \Big] \nonumber\\\\[1.2ex]
 	&& \hspace{0.9cm}\cdot \Pr \Big[S^n(1,{\ell'})=s'^n \big| \nonumber\\&& \qquad\qquad\;\;\;  (S^n(1,1),Y^n)\in \mathcal{T}_{\mu}^n(P_{SY}),\;\;\nonumber\\&& \qquad\qquad\;\;\;  (S^n(1,1),X^n)\in \mathcal{T}_{\mu/2}^n(P_{SX}),\nonumber\\&& \qquad\qquad\;\;\; S^n(1,{\ell'}) \in\mathcal{T}_{\mu}^n(P_S)),\;M=L=1,\; \mathcal{H}=0\Big] \nonumber \\\\[1.2ex]
 	&\stackrel{(e)}{\leq} & (n+1)^{|\mathcal{S}|^2\cdot |\mathcal{Y}|} \sum_{\pi_{SS'Y}\in\mathcal{P}_{\mu,\text{type-I}}^{n}}\sum_{{\ell'}=2}^{\lfloor 2^{nR'}\rfloor}\sum_{\substack{s^n,y^n,s'^n:\\\text{tp}(s^n,s'^{n},y^n)=\pi_{SS'Y}}} \nonumber\\&& \hspace{2cm}2^{-nH_{\pi}(S,Y)}\cdot 2^{-nH_{\pi}(S')}  \\[1ex]
 	&\stackrel{(f)}{\leq}& (n+1)^{|\mathcal{S}|^2\cdot |\mathcal{Y}|} \sum_{\pi_{SS'Y}\in\mathcal{P}_{\mu,\text{type-I}}^{n}}\sum_{{\ell'}=2}^{\lfloor 2^{nR'}\rfloor}\nonumber\\&& \hspace{1.7cm} 2^{nH_{\pi}(S,S',Y)}\cdot 2^{-nH_{\pi}(S,Y)}\cdot 2^{-nH_{\pi}(S')}\\[1ex]
 	&=&(n+1)^{|\mathcal{S}|^2\cdot |\mathcal{Y}|} \sum_{\pi_{SS'Y}\in\mathcal{P}_{\mu,\text{type-I}}^{n}}  2^{n(R'-I_{\pi}(S';Y,S))}  \\[0.8ex]
 	&\leq&(n+1)^{|\mathcal{S}|^2\cdot |\mathcal{Y}|} \sum_{\pi_{SS'Y}\in\mathcal{P}_{\mu,\text{type-I}}^{n}}  2^{n(R'-I_{\pi}(S';Y))}  \\[0.8ex]
 	&\stackrel{(g)}{\leq} &  (n+1)^{|\mathcal{S}|^4\cdot |\mathcal{Y}|^2}\cdot \max_{\pi_{SS'Y}\in\mathcal{P}_{\mu,\text{type-I}}^{n}}  2^{n(R'-I(S;Y)+\delta_n(\mu))} \label{eq:lastss}  \\
 	& \stackrel{(h)}{\leq} & \epsilon/6,
 \end{IEEEeqnarray}
 where $\delta_n(\mu)$ tends to $0$ as $n \to \infty$ and then $\mu\to 0$. The steps are justified as follows:
 \begin{itemize}
	\item $(d)$ holds because conditioned on  the events $(S^n(1,1),Y^n)\in \mathcal{T}_{\mu}^n(P_{SY}),\;\;(S^n(1,1),X^n)\in \mathcal{T}_{\mu/2}^n(P_{SX}),\; \; S^n(1,{\ell'}) \in\mathcal{T}_{\mu}^n(P_S) ,\;\;M=L=1,\;\; \mathcal{H}=0$, the codeword $S^n(1,\ell')$ is  independent of the pair $(S^n(1,1),Y^n)$;
 	\item $(e)$  holds because  even conditioned on the events $(S^n(1,1),Y^n)\in \mathcal{T}_{\mu}^n(P_{SY}),\;\;(S^n(1,1),X^n)\in \mathcal{T}_{\mu/2}^n(P_{SX}),\; \; S^n(1,{\ell'}) \in\mathcal{T}_{\mu}^n(P_S)) ,\;\;M=L=1,\;\; \mathcal{H}=0$, all pairs $(s^n,y^n)$ of same joint type have the same probability  and all sequences $s'^{n}$ of same type have the same probability, and because there are at least $\frac{1}{(n+1)^{|\mathcal{S}|\cdot |\mathcal{Y}|}}\cdot 2^{nH_{\pi_{SY}}(S,Y)}$ sequences of joint type $\pi_{SY}$ \cite[Lemma~2.3]{Csiszarbook}  and at least  $\frac{1}{(n+1)^{|\mathcal{S}|}} \cdot 2^{nH_{\pi_{S'}}(S')}$ sequences of joint type $\pi_{S'}$;
 	\item
 	$(f)$ {because  there are at most $2^{nH_{\pi}(S,S',Y)}$ different $n$-length sequences of same joint type $\pi_{SS'Y}$;}
	 	\item $(g)$  holds because $|\mathcal{P}_{\mu,\text{type-I}}^{n}| \leq (n+1)^{|\mathcal{S}|^2 \cdot |\mathcal{Y}|}$, because $H_{\pi}(S'|Y) \leq H_{\pi}(S|Y)$, because each element of $\mathcal{P}_{\mu,\text{type-I}}^n$ must approach an element of $\mathcal{P}_{\text{type-I}}^*$ when $n\to\infty$ and $\mu\to0$, and  by the continuity of the entropy function; and
 	\item $(h)$ holds for all sufficiently large $n$ and small $\mu$   because    $R'< I(S;Y)$ and $\delta_n(\mu) \to 0$ as  $n\to \infty$ and then $\mu \to 0$.
 \end{itemize}

We now bound the probability of type-II error at the receiver (averaged over the random code construction).
For all $m\in \{1,\ldots, \lfloor 2^{nR}\rfloor \}$ and $\ell,\ell'\in \{1,\ldots, \lfloor 2^{nR'}\rfloor \}$, define the following events:
\begin{IEEEeqnarray}{rCl}
	\mathcal{E}_{\text{Tx}}(m,\ell)&:= & \{(S^n(m,\ell),X^n)\in \mathcal{T}_{\mu/2}^n(P_{SX})\},\\
	\mathcal{E}_{\text{Rx}}(m,\ell')&:= & \Big\{(S^n(m,\ell'),Y^n)\in \mathcal{T}_{\mu}^n(P_{SY}),\nonumber\\&&\hspace{-1.2cm} H_{\text{tp}(S^n(m,\ell'),Y^n)}(S'|Y)=\min_{\tilde{\ell}}H_{\text{tp}(S^n(m,\tilde{\ell}),Y^n)}(S|Y)\Big\}.\nonumber\\
	\end{IEEEeqnarray}
Define
\begin{IEEEeqnarray}{rCl}
	\mathcal{B}_1&:=& \{ \exists\; (m,\ell)\colon \mathcal{E}_{\text{Tx}}(m,\ell) \text{ and } \mathcal{E}_{\text{Rx}}(m,\ell) \},\\
		\mathcal{B}_2&:=& \{ \exists\; (m,\ell,\ell'), \ell\neq \ell'\colon \mathcal{E}_{\text{Tx}}(m,\ell) \text{ and } \mathcal{E}_{\text{Rx}}(m,\ell') \}.\nonumber\\
	\end{IEEEeqnarray}
Then we have:
\begin{align}
\mathbb{E}_{\mathcal{C}}[\beta_{y,n}] \leq \sum_{i=1}^2 \Pr \left[ \mathcal{B}_i \big| \mathcal{H}=1 \right].\label{type-II}
\end{align}
We bound each of the probabilities on the right-hand side of \eqref{type-II}. We introduce the following type classes:
\begin{IEEEeqnarray}{rCl}
\mathcal{P}_{\mu, 1} &:=& \left\{ \pi_{SXY}\colon |\pi_{SX}-P_{SX}|<\mu/2,  \ \   |\pi_{SY}-P_{SY}|<\mu \right\},\nonumber\\\\
\mathcal{P}_{\mu, 2} &:=& \Big\{ \pi_{SS'XY}\colon |\pi_{SX}-P_{SX}|<\mu/2,  \nonumber\\*&& \hspace{-3mm} |\pi_{S'Y}-P_{SY}|<\mu,\quad H_{\pi}(S'|Y)\leq H_{\pi}(S|Y) \Big\}.
	\end{IEEEeqnarray}
Consider $\mathcal{B}_1$ as follows:
\begin{IEEEeqnarray}{rCl}
\Pr\left[ \mathcal{B}_1|\mathcal{H}=1 \right]
	&\leq& \sum_{m,\ell} \;\;\Pr\left[\mathcal{E}_{\text{Tx}}(m,\ell)\;\cap\; \mathcal{E}_{\text{Rx}}(m,\ell)\big|\mathcal{H}=1 \right]\nonumber\\
	&\leq &\sum_{m,\ell}\;\; \Pr \Big[ (S^n(m,\ell),X^n)\in \mathcal{T}_{\mu/2}^n(P_{SX}),\nonumber\\&&\hspace{1.1cm}(S^n(m,\ell),Y^n)\in \mathcal{T}_{\mu}^n(P_{SY}) \big|\mathcal{H}=1 \Big]\nonumber\\
	&=&\sum_{m,\ell}\;\;\;\sum_{\substack{\pi_{SXY}:\\ |\pi_{SX}-P_{SX}|<\mu/2,\\|\pi_{SY}-P_{SY}|<\mu}}\nonumber\\&&\Pr\left[\text{tp}\left( S^n(m,\ell),X^n,Y^n\right)=\pi_{SXY} \big| \mathcal{H}=1 \right]\nonumber\\[2.5ex]
	&\leq & 2^{n(R+R')}\cdot \;\; (n+1)^{|\mathcal{S}|\cdot|\mathcal{X}|\cdot|\mathcal{Y}|}\nonumber\\&&\cdot\max_{\substack{\pi_{SXY}:\\ |\pi_{SX}-P_{SX}|<\mu/2,\\|\pi_{SY}-P_{SY}|<\mu}} 2^{-n(D(\pi_{SXY}|| P_SQ_{XY})-\mu)},\nonumber\\\label{t1-bound}
	\end{IEEEeqnarray}
where the last inequality follows from Sanov's theorem and the i.i.d. codebook construction. Define now:
\begin{align}
\tilde{\theta}_{\mu,1}:{=} \min_{\substack{\pi_{SXY}:\\ |\pi_{SX}-P_{SX}|<\mu/2,\\|\pi_{SY}-P_{SY}|<\mu}} D(\pi_{SXY}||P_SQ_{XY})-R-R'-\mu,
\end{align}
and notice that
\begin{IEEEeqnarray}{rCl}
\lefteqn{\tilde{\theta}_{\mu,1}}~~\nonumber\\
  &\stackrel{\text{Eq.}\;\eqref{cas1binp2pb}}{=}&\!\!\!\!\!\!\min_{\substack{\pi_{SXY}:\\|\pi_{SX}-P_{SX}|<\mu/2,\\|\pi_{SY}-P_{SY}|<\mu}} D(\pi_{SXY}||P_SQ_{XY})-I(S;X)-2\mu\nonumber\\
&=&\!\!\!\!\!\! \min_{\pi_{SXY}\in\mathcal{P}_{\mu,1}} \sum_{s,x,y} \Bigg[\pi_{SXY}(s,x,y)\log \frac{ \pi_{SXY}(s,x,y)}{P_S(s)Q_{XY}(x,y)}\nonumber\\&&\hspace{2cm}-P_{SX}(s,x)\log\frac{P_{S|X}(s,x)}{P_{S}(s)}\Bigg]-2\mu\nonumber\\
&\stackrel{(j)}{=}&\!\!\!\!\!\!\!\! \min_{\pi_{SXY}\in\mathcal{P}_{\mu,1}} \sum_{s,x,y} \Bigg[\pi_{SXY}(s,x,y)\log \frac{ \pi_{SXY}(s,x,y)}{P_S(s)Q_{XY}(x,y)}\nonumber\\&&\hspace{2cm}-\pi_{SX}(s,x)\log\frac{P_{S|X}(s,x)}{P_{S}(s)}\Bigg]-\delta_1(\mu)\nonumber\\
&\stackrel{(k)}{=}&\!\!\!\!\!\!\!\! \min_{\pi_{SXY}\in\mathcal{P}_{\mu,1}} \!\sum_{s,x,y} \pi_{SXY}(s,x,y)\log \frac{ \pi_{SXY}(s,x,y)}{P_{S|X}(s|x)Q_{XY}(x,y)}\nonumber\\&&\hspace{2.3cm}-\delta_1(\mu)\nonumber\\[1.5ex]
&=&\!\!\!\!\!\! \min_{\pi_{SXY}\in\mathcal{P}_{\mu,1}} D(\pi_{SXY}\|P_{S|X}Q_{XY})-\delta_1(\mu)\nonumber\\
&{=}& \theta_{\mu,1}-\delta_1(\mu),\label{t1-final}
\end{IEEEeqnarray}
for a function $\delta_1(\mu)$ that goes to zero as $\mu\to 0$ and 
\begin{IEEEeqnarray}{rCl}
	 \theta_{\mu,1} := \min_{\pi_{SXY}\in\mathcal{P}_{\mu,1}} D(\pi_{SXY}\|P_{S|X}Q_{XY}).
	\end{IEEEeqnarray} Here, $(j)$ holds because  $|\pi_{SX}-P_{SX}|<\mu/2$ and by the continuity of the KL-divergence; $(k)$ follows by re-arranging terms. Considering \eqref{t1-bound} and \eqref{t1-final} yields the following:
\begin{align}
\Pr\left[ \mathcal{B}_1|\mathcal{H}=1 \right] \leq (n+1)^{|\mathcal{S}|\cdot|\mathcal{X}|\cdot|\mathcal{Y}|}\cdot \; 2^{-n(\theta_{\mu,1}-\delta_1(\mu))}.\label{b1-event}
\end{align}

Next, consider $\mathcal{B}_2$ as follows:
\begin{IEEEeqnarray}{rCl}
	&&\hspace{-0.5cm}\Pr \left[ \mathcal{B}_2 \big| \mathcal{H}=1 \right] \nonumber\\&\leq & \sum_{m}\;\;\sum_{\ell\neq\ell'} \;\;\Pr\left[\mathcal{E}_{\text{Tx}}(m,\ell)\;\cap\; \mathcal{E}_{\text{Rx}}(m,\ell')\big|\mathcal{H}=1 \right]\nonumber\\
	&=& \sum_{m}\;\;\sum_{\ell\neq\ell'} \;\; \Pr \Bigg[ (S^n(m,\ell),X^n)\in \mathcal{T}_{\mu/2}^n(P_{SX}),\nonumber\\&&\hspace{2.3cm}(S^n(m,\ell'),Y^n)\in \mathcal{T}_{\mu}^n(P_{SY}),\nonumber\\&&\hspace{2.3cm} H_{\text{tp}(S^n(m,\ell'),Y^n)}(S'|Y)\nonumber\\[-1ex]&&\hspace{2.5cm}=\min_{\tilde{\ell}}H_{\text{tp}(S^n(m,\tilde{\ell}),Y^n)}(S|Y) \Big|\mathcal{H}=1 \Bigg]\nonumber\\
	&=&\sum_{m}\;\;\sum_{\ell\neq\ell'}\;\;\sum_{\substack{\pi_{SS'XY}:\\ |\pi_{SX}-P_{SX}|<\mu/2,\\|\pi_{S'Y}-P_{SY}|<\mu\\H_{\pi}(S'|Y)\leq H_{\pi}(S|Y)}}\nonumber\\&&\Pr\left[\text{tp}\left( S^n(m,\ell),S^n(m,\ell'),X^n,Y^n\right)=\pi_{SS'XY} \big| \mathcal{H}=1 \right]\nonumber\\[2ex]
	&\leq & 2^{n(R+2R')}\cdot \;\; (n+1)^{|\mathcal{S}|^2\cdot|\mathcal{X}|\cdot|\mathcal{Y}|} \nonumber\\&&\hspace{0.5cm}\cdot\hspace{-3mm}\max_{\substack{\pi_{SS'XY}:\\ |\pi_{SX}-P_{SX}|<\mu/2,\\|\pi_{S'Y}-P_{SY}|<\mu\\H_{\pi}(S'|Y)\leq H_{\pi}(S|Y)}}2^{-n(D(\pi_{SS'XY}||P_SP_SQ_{XY})-\mu)},\label{t2-bound}
	\end{IEEEeqnarray}
where the last inequality follows from Sanov's theorem. 
Now, define:
\begin{align}
\tilde{\theta}_{\mu,2} &:{=}\hspace{-0.3cm} \min_{\substack{\pi_{SS'XY}:\\ |\pi_{SX}-P_{SX}|<\mu/2,\\|\pi_{S'Y}-P_{SY}|<\mu\\H_{\pi}(S'|Y)\leq H_{\pi}(S|Y)}}\hspace{-0.5cm} D(\pi_{SS'XY}||P_SP_SQ_{XY})-R-2R'-\mu.
\end{align}
Consider the following chain of inequalities:
\begin{IEEEeqnarray}{rCl}
	\tilde{\theta}_{\mu,2} &\stackrel{\text{Eq.}\;\eqref{cas1binp2pb}}{=}& \hspace{-0.5cm} \min_{\substack{\pi_{SS'XY}:\\ |\pi_{SX}-P_{SX}|<\mu/2,\\|\pi_{S'Y}-P_{SY}|<\mu\\H_{\pi}(S'|Y)\leq H_{\pi}(S|Y)}}\hspace{-0.4cm} D(\pi_{SS'XY}||P_SP_SQ_{XY})\nonumber\\[-2ex]&&\hspace{2.5cm}-2I(S;X)+R-3\mu\nonumber\\[1.5ex]
	&=&\min_{\substack{\pi_{SS'XY}\in\mathcal{P}_{\mu,2}^n}} D(\pi_{SS'XY}||P_SP_SQ_{XY})\nonumber\\[-0.5ex]&&\hspace{2.5cm}-2I(S;X)+R-3\mu\nonumber\\[1ex]
	&\stackrel{(l)}{=}&  \min_{\substack{\pi_{SS'XY}\in\mathcal{P}_{\mu,2}^n}} \Big[D(\pi_{SXY}||P_SQ_{XY})\nonumber\\[-1ex]&&\hspace{2.2cm}+\mathbb{E}_{\pi_{SXY}}\left[ D(\pi_{S'|SXY}\|P_S)\right]\Big]\nonumber\\&&\hspace{2.2cm}-2I(S;X)+R-3\mu\nonumber\\[1ex]
	&\stackrel{(m)}{\geq} &  \min_{\substack{\pi_{SS'XY}\in\mathcal{P}_{\mu,2}^n}} \Big[D(\pi_{SXY}||P_SQ_{XY})\nonumber\\[-1ex]&&\hspace{2.2cm}+\mathbb{E}_{\pi_{Y}}\left[ D(\pi_{S'|Y}\|P_S)\right]\Big]\nonumber\\&&\hspace{2.2cm}-2I(S;X)+R-3\mu\nonumber\\[1ex]
	&\stackrel{(n)}{=}&\min_{\substack{\pi_{SS'XY}\in\mathcal{P}_{\mu,2}^n}} D(\pi_{SXY}||P_SQ_{XY})\nonumber\\[0ex]&&\hspace{1.7cm}+I(S;Y)-2I(S;X)+R-\delta_2'(\mu)\nonumber\\[1ex]
	&\stackrel{(o)}{=}&\min_{\substack{\pi_{SS'XY}\in\mathcal{P}_{\mu,2}^n}} D(\pi_{SXY}||P_{S|X}Q_{XY})\nonumber\\[0ex]&&\hspace{1.7cm}+I(S;Y)-I(S;X)+R-\delta_2(\mu)\nonumber\\[1ex]
	&=& \theta_{\mu,2}-\delta_2(\mu),\label{t2-final}
	\end{IEEEeqnarray}
for   functions $\delta'_2(\mu), \delta_2(\mu)$ that go to zero as $\mu\to 0$ and 
\begin{IEEEeqnarray}{rCl}
\theta_{\mu,2}&:=& 	\min_{\substack{\pi_{SS'XY}\in\mathcal{P}_{\mu,2}^n}} D(\pi_{SXY}||P_{S|X}Q_{XY})\nonumber\\[0ex]&&\hspace{1.8cm}+I(S;Y)-I(S;X)+R.
	\end{IEEEeqnarray}
		 Here, $(l)$ follows from the chain rule for KL-divergence; $(m)$ follows from the convexity of the KL-divergence and Jensen's inequality; $(n)$ follows because $|\pi_{S'Y}-P_{SY}|<\mu$ and by the continuity of  KL-divergence; $(o)$ follows by re-arranging terms and employing similar steps leading to \eqref{t1-final}.
Combining \eqref{t2-bound} and \eqref{t2-final} yields the following:

\begin{align}
\Pr\left[ \mathcal{B}_2|\mathcal{H}=1 \right] \leq (n+1)^{|\mathcal{S}|^2\cdot|\mathcal{X}|\cdot|\mathcal{Y}|}\cdot \; 2^{-n(\theta_{\mu,2}-\delta_2(\mu))}.\label{b2-event}
\end{align} 

Combining \eqref{type-II}, \eqref{b1-event}, and \eqref{b2-event} proves that for large blocklengths $n$:
\begin{align}
\mathbb{E}_{\mathcal{C}}\left[ \beta_{y,n}\right] &\leq (n+1)^{|\mathcal{S}|\cdot|\mathcal{X}|\cdot|\mathcal{Y}|}\cdot \; 2^{-n(\theta_{\mu,1}-\delta_1(\mu))} \nonumber\\&\hspace{0.5cm}+ (n+1)^{|\mathcal{S}|^2\cdot|\mathcal{X}|\cdot|\mathcal{Y}|}\cdot \; 2^{-n(\theta_{\mu,2}-\delta_2(\mu))}.
\end{align}
Letting $n\to \infty$ and then $\mu\to 0$, we get that $\theta_{\mu,1}\to \theta_1$ and $\theta_{\mu,2}\to \theta_2$, where we define:
\begin{subequations}
	\label{eq:binning_exponents}
\begin{IEEEeqnarray}{rCl}
	\theta_1 &:=& \hspace{-3mm} \min_{\substack{\tilde{P}_{SXY}:\\\tilde{P}_{SX}=P_{SX}\\\tilde{P}_{SY}=P_{SY}}} D(\tilde{P}_{SXY}\|P_{S|X}Q_{XY}), \\
	\theta_2 &:= & \hspace{-0.8cm} \min_{\substack{\tilde{P}_{SXY}:\\\tilde{P}_{SX}=P_{SX}\\\tilde{P}_{Y}=P_{Y}\\ H(S|Y)\leq H_{\tilde{P}_{SY}}(S|Y)\\[2ex]}} \hspace{-0.9cm}D(\tilde{P}_{SXY}\|P_{S|X}Q_{XY})\nonumber\\*
	& & \hspace{1cm} {}+R -I(S;X)+I(S;Y),
	\end{IEEEeqnarray}
\end{subequations}
where $P_{SY}$ in the minimization constraint is  the marginal pmf of $P_{SXY}=P_{S|X} P_{XY}$ and the conditional entropy term  $H(S|Y)$  is calculated according to this marginal.

The theorem  then follows immediately by \eqref{eq:binning_exponents} and  $I(S;X)-I(S;Y)=I(S;X|Y)$ (which holds by the Markov chain $S - X - Y$), and from the fact that $P_{S|X}$ can be chosen arbitrary.

\section{Proof of Theorem~\ref{thm:casbin}}\label{casbinpr}
We analyze the probabilities of error of the coding and testing scheme described in Subsection~\ref{sec:cascade_binning} averaged over the random code constructions. By successively eliminating the worst half of  the codebooks, as sketched for example at the end of Appendix~\ref{cascadeachnbpr}), the desired result can be proved for a set of deterministic codebooks.

Fix an arbitrary $\epsilon>0$ and the parameter of the scheme $\mu$ sufficiently close to $0$ as will become clear in the sequel. Fix also a blocklength $n$. 
 If $M\neq 0$, let $I,J$ be the indices sent from the transmitter to the relay. If both $B\neq 0$ and $M\neq 0$, let $K$ denote the second index sent from the relay to the receiver. 

We first analyze the type-I error probability at the receiver. 
Define events:
\begin{align}
\mathcal{E}_{\text{Rx}}^{(1)}&\colon \Big\{ \exists  f' \neq F\colon  H_{\text{tp}(S^n(I),V^n(K,f'|I),Z^n)}(V|S,Z)\nonumber\\*&\hspace{2cm}= \min_{\tilde{f} } H_{\text{tp}(S^n(I),V^n(K,\tilde{f}|I),Z^n)}(V|S,Z)\Big\},\\
\mathcal{E}_{\text{Rx}}^{(2)} &\colon \left\{ (X^n(I),V^n(K,F|I),Z^n)\notin \mathcal{T}_{\mu}^n(P_{SVZ})\right\}.
\end{align}
The type-I error probability can then be bounded as follows:
\begin{align}
\mathbb{E}_{\mathcal{C}}[\alpha_{z,n}] &\leq \Pr\left[M=0\;\cup\;B=0\;\cup\;
\mathcal{E}_{\text{Rx}}^{(1)}\;\cup\;\mathcal{E}_{\text{Rx}}^{(2)}\right]\nonumber\\*
&\leq \Pr\left[M=0\right] +
\Pr\left[B=0|M\neq 0\right]\nonumber\\&\hspace{0.5cm}+\Pr\left[\mathcal{E}_{\text{Rx}}^{(1)}\big|M\neq 0, B\neq 0\right]\nonumber\\&\hspace{0.5cm}+\Pr\left[\mathcal{E}_{\text{Rx}}^{(2)}\big|M\neq 0, B\neq 0,\mathcal{E}_{\text{Rx}}^{(1)c}\right]\nonumber\\
&\stackrel{(a)}{\leq} \epsilon/16+\Pr\left[B=0|M\neq 0\right]\nonumber\\&\hspace{0.5cm}+\Pr\left[\mathcal{E}_{\text{Rx}}^{(2)}\big|M\neq 0, B\neq 0\right]\nonumber\\&\hspace{0.5cm}+\Pr\left[\mathcal{E}_{\text{Rx}}^{(1)}\big|M\neq 0, B\neq 0,\mathcal{E}_{\text{Rx}}^{(1)c}\right]\nonumber\\
&\stackrel{(b)}{\leq}\epsilon/16+\epsilon/16+\Pr\left[\mathcal{E}_{\text{Rx}}^{(2)}\big|M\neq 0, B\neq 0\right]\nonumber\\&\hspace{0.5cm}+\Pr\left[\mathcal{E}_{\text{Rx}}^{(1)}\big|M\neq 0, B\neq 0,\mathcal{E}_{\text{Rx}}^{(1)c}\right]\nonumber\\
&\stackrel{(c)}{\leq}\epsilon/16+\epsilon/16+\epsilon/16\nonumber\\&\hspace{0.5cm}+\Pr\left[\mathcal{E}_{\text{Rx}}^{(1)}\big|M\neq 0, B\neq 0,\mathcal{E}_{\text{Rx}}^{(2)c}\right]\nonumber\\
&\stackrel{(d)}{\leq}\epsilon/4.
\end{align}
where $(a)$ holds by the covering lemma and the rate-constraints in \eqref{eq:rates_cascd}; $(b)$ and $(d)$ can be proved following similar lines as the type-I error analysis in Appendix~\ref{thm1pr}; and $(c)$  holds by the Markov lemma.

We now bound the probability of type-II error at the receiver. Define the following events:
\begin{IEEEeqnarray}{rCl}
	&&\mathcal{E}_{\text{Tx}}(i,j,e) =\left\{ (S^n(i),U^n(j,e|i),X^n)\in \mathcal{T}_{\mu/4}^n(P_{SUX}) \right\},\nonumber\\* \ \\
	&&\mathcal{E}_{\text{Rel}}(i,j,e',k,f) =\nonumber\\&&\hspace{0.8cm} \Big\{\left(S^n(i),U^n(j,e'|i),V^n(k,f|i),Y^n\right)\in \mathcal{T}_{\mu/2}^n(P_{SUVY}),\nonumber\\&&\hspace{1.5cm} H_{\text{tp}\left(S^n(i),U^n(j,e'|i),Y^n\right)}(U|S,Y)=\nonumber\\&&\hspace{3cm}\min_{\tilde{e}} H_{\text{tp}\left(S^n(i),U^n(j,\tilde{e}|i),Y^n\right)}(U|S,Y)\Big\},\nonumber\\\\
	&&\mathcal{E}_{\text{Rx}}(i,k,f') =\nonumber\\&&\hspace{0.8cm} \Big\{ (S^n(i),V^n(k,f'|i),Z^n)\in \mathcal{T}_{\mu}^n(P_{SVZ}),\nonumber\\&&\hspace{1.5cm} H_{\text{tp}\left(S^n(i),V^n(k,f'|i),Z^n\right)}(V|S,Z)=\nonumber\\&&\hspace{3cm}\min_{\tilde{f}} H_{\text{tp}\left(S^n(i),V^n(k,\tilde{f}|i),Z^n\right)}(V|S,Z)\Big\}.\nonumber\\
	\end{IEEEeqnarray}
We then have: 
\begin{IEEEeqnarray}{rCl}
	\mathbb{E}_{\mathcal{C}}\left[\beta_{z,n} \right] & =& \Pr \left[ \hat{\mathcal{H}}_z=0 \Big| \mathcal{H}=1 \right] \\
	&  \leq 
& \Pr\left[ B \neq 0 , \  \bigcup_{i,k,f'}\; \mathcal{E}_{\text{Rx}}(i,k,f') \Big|\mathcal{H}=1 \right]\nonumber\\\\
& \leq &  \Pr\Big[ \bigcup_{i,j,e,e',k,f,f'}\; \mathcal{E}_{\text{Tx}}(i,j,e)    \nonumber\\&&\hspace{2cm} \textnormal{ and }  \mathcal{E}_{\text{Rel}}(i,j,e',k,f)\nonumber\\&&\hspace{2cm}\textnormal{ and }\;\mathcal{E}_{\text{Rx}}(i,k,f') \Big|\mathcal{H}=1 \Big]\nonumber\\
	\end{IEEEeqnarray}
We can further upper bound this last probability with the union bound to obtain: 
\begin{IEEEeqnarray}{rCl}
	\mathbb{E}_{\mathcal{C}}\left[ \beta_{z,n}\right] \leq \sum_{i=1}^{4}\Pr\left[\mathcal{B}_i\big| \mathcal{H}=1 \right],\label{20-sum}
	\end{IEEEeqnarray}
where the four events $\mathcal{B}_1,\mathcal{B}_2,\mathcal{B}_3, \mathcal{B}_4$ are defined as:
\begin{IEEEeqnarray}{rCl}
	\mathcal{B}_1&\colon& \Big\{ \exists\; (i,j,e,k,f)\colon \;\;\mathcal{E}_{\text{Tx}}(i,j,e)\;\;\text{and} \;\;   \nonumber\\&&\hspace{2.7cm} \mathcal{E}_{\text{Rel}}(i,j,e,k,f)\;\;\text{and}\; \mathcal{E}_{\text{Rx}}(i,k,f) \Big\},\nonumber\\\\
				\mathcal{B}_2&\colon& \Big\{ \exists \;(i,j,e,e',k,f)\colon\;\; e\neq e'\;\;\text{and}\;\;  \mathcal{E}_{\text{Tx}}(i,j,e)\;\; \text{and} \;\; \nonumber\\&&\hspace{2.7cm}  \mathcal{E}_{\text{Rel}}(i,j,e',k,f)\;\;\text{and}\;\; \mathcal{E}_{\text{Rx}}(i,k,f) \Big\},\nonumber\\\\
		\mathcal{B}_3&\colon& \Big\{ \exists\; (i,j,e,k,f,f')\colon\;\; f\neq f'\;\;\text{and}\;\;  \mathcal{E}_{\text{Tx}}(i,j,e)\;\;  \text{and}\;\; \nonumber\\&&\hspace{2.7cm} \mathcal{E}_{\text{Rel}}(i,j,e,k,f)\;\; \text{and}\;\; \mathcal{E}_{\text{Rx}}(i,k,f') \Big\},\nonumber\\\\
			\mathcal{B}_4&\colon& \Big\{ \exists\; (i,j,e,e',k,f,f')\colon\;\; e\neq e'\;\;\text{and}\;\; f\neq f'\;\; \text{and}\;\;\nonumber\\&&\hspace{0.5cm}\mathcal{E}_{\text{Tx}}(i,j,e)\;\; \text{and} \;\;  \mathcal{E}_{\text{Rel}}(i,j,e',k,f)\;\; \text{and}\;\; \mathcal{E}_{\text{Rx}}(i,k,f') \Big\}.\nonumber\\
	\end{IEEEeqnarray}
The summands in \eqref{20-sum} can be analyzed by now standard arguments as used in Appendices~\ref{cascadeachnbpr} and \ref{thm1pr}.  
	 
For each $i=1,2,3,4$, this yields an exponential bound of the form
\begin{equation}\label{eq:exp}
\Pr[\mathcal{B}_i] \leq 2^{-n( \theta_{i} + \delta_i(\mu))},
\end{equation}
where $ \delta_{1}(\mu),  \delta_{2}(\mu),  \delta_{3}(\mu), \delta_{4}(\mu)$ are functions that tend to 0 as $\mu \to 0$ and where
\begin{align}
&\theta_{1} :{=} \min_{\substack{\pi_{SUVXYZ} \colon\\ \pi_{SUX}=P_{SUX} \\ \pi_{SUVY}=P_{SUVY}\\ \pi_{SVZ} =P_{SVZ} }} D(\pi_{SUVXYZ}\|P_{US|X}P_{V|SUY}Q_{XYZ}),\label{t0}\\[1ex]
&\theta_{2} :{=}\min_{\substack{\pi_{SUU'VXYZ} \colon \\\pi_{SUX}=P_{SUX} \\
		H(U|S,Y) \leq H_{\pi}(U|S,Y) \\ \pi_{SU'VY}=P_{SU'VY}\\ \pi_{SVZ} =P_{SVZ} }}\nonumber\\&\hspace{1.8cm}D(\pi_{SUU'VXYZ}\| P_{SU|X}P_{U'|S}P_{V|SU'Y}Q_{XYZ}) \nonumber\\&\hspace{2.5cm}+R_u - I(U;X|S), \label{t3}\\[1ex]
&\theta_{3} :{=} \min_{\substack{\pi_{SUVV'XYZ} \colon\\ \pi_{SUX}=P_{SUX} \\ \pi_{SUVY}=P_{SUVY}\\ H(V|S,Z) \leq H_{\pi}(V|S,Z) \\\pi_{SV'Z} =P_{SV'Z} }}\nonumber\\& \hspace{2cm} D(\pi_{SUVV'XYZ}\|P_{SU|X}P_{V|SUY}P_{V'|S}Q_{XYZ})\nonumber\\&\hspace{2.5cm}+ R_v-I(V;U,Y|S),\label{t1}\\[1ex]
&\theta_{4} :{=}  \min_{\substack{\pi_{SUU'VV'XYZ} \colon\\ \pi_{SUX}=P_{SUX} \\ H(U|S,Y) \leq H_{\pi}(U|S,Y)  \\ \pi_{SU'VY}=P_{SU'VY}\\ H(V|S,Z) \leq H_{\pi}(V|S,Z) \\\pi_{SV'Z} =P_{SV'Z} }}\nonumber\\&\hspace{0.7cm} D(\pi_{SUU'VV'XYZ}\|P_{SU|X}P_{U'|S}P_{V|SU'Y} P_{V'|S}Q_{XYZ})\nonumber\\&\hspace{1.5cm}+R_u+R_v- I(U;X|S) - I(V;U,Y|S).\label{t2}
\end{align}
Plugging the exponential bounds \eqref{eq:exp} into \eqref{20-sum}, extracting the term $I(U';Y|S)=I(U;Y|S)$ from \eqref{t3} and \eqref{t2} and the term $I(V';Z|S)=I(V;Z|S)$ from \eqref{t1} and \eqref{t2},  and bounding $R_u$ and $R_v$ by $R- I(S;X)$ and $T-I(S;X)$, we obtain the result in the theorem.

\section{Proof of Proposition~\ref{prop:Markov}}\label{app:proof_prop}

The inclusion 
\begin{equation}
\mathcal{E}_{\simple}(R,T) \subseteq \mathcal{E}_{\nobinning}(R,T),
\end{equation} 
is straightforward. It suffices to
note that  restricting the union in \eqref{eq:Eidef} to choices of the conditional pmfs $P_{SU|X}$ and $P_{V|SUY}$ where $S$ is a constant and $V$ is conditionally independent of $U$ given $Y$,  results in $\mathcal{E}_{\simple}(R,T)$.  

We now prove the reverse inclusion
\begin{equation}\label{eq:inc2}
\mathcal{E}_{\simple}(R,T) \supseteq \mathcal{E}_{\nobinning}(R,T).
\end{equation}
Fix an arbitrary pair $P_{SU|X}$ and $P_{V|SUY}$  satisfying the rate-constraints \eqref{casach3}--\eqref{casach4}.
Then, notice the following sequence of equalities:
\begin{align}\lefteqn{
	\min_{\substack{\tilde{P}_{SUVXYZ}:\\\tilde{P}_{SUX}=P_{SUX}\\\tilde{P}_{SUVY}=P_{SUVY}\\\tilde{P}_{SVZ}=P_{SVZ}}} D(\tilde{P}_{SUVXYZ}\|P_{SU|X}P_{V|SUY}Q_{XY}Q_{Z|Y})}\nonumber \qquad \\
&= \min_{\substack{\tilde{P}_{SUXY}:\\\tilde{P}_{SUX}=P_{SUX}\\\tilde{P}_{SUY}=P_{SUY}}} \bigg[ D(\tilde{P}_{SUXY}\|P_{SU|X}Q_{XY})\nonumber\\&\hspace{0.05cm}+  \mathbb{E}_{\tilde{P}_{SUXY}}\Big[\hspace{-0.3cm} \min_{\substack{\tilde{P}_{VZ|SUXY}:\\\tilde{P}_{V|SUY}=P_{V|SUY}\\\tilde{P}_{Z|SV}=P_{Z|SV}}} \hspace{-0.5cm} D(\tilde{P}_{VZ|SUXY}\|P_{V|SUY}Q_{Z|Y}) \Big] \bigg]\nonumber\\
&\stackrel{(a)}{=} \min_{\substack{\tilde{P}_{SUXY}:\\\tilde{P}_{SUX}=P_{SUX}\\\tilde{P}_{SUY}=P_{SUY}}} \bigg[ D(\tilde{P}_{SUXY}\|P_{SU|X}Q_{XY})\nonumber\\&\hspace{0.3cm}+  \mathbb{E}_{{P}_{SUY}}\Big[\hspace{-0.3cm} \min_{\substack{\tilde{P}_{VZ|SUY}:\\\tilde{P}_{V|SUY}=P_{V|SUY}\\\tilde{P}_{Z|SV}=P_{Z|SV}}}  \hspace{-0.5cm}D(\tilde{P}_{VZ|SUY}\|P_{V|SUY}Q_{Z|Y}) \Big] \bigg]\nonumber\\
&\stackrel{(b)}= \min_{\substack{\tilde{P}_{SUXY}:\\\tilde{P}_{SUX}=P_{SUX}\\\tilde{P}_{SUY}=P_{SUY}}} \bigg[ D(\tilde{P}_{SUXY}\|P_{SU|X}Q_{XY}) \nonumber\\&\hspace{0.5cm}+\mathbb{E}_{{P}_{SUVY}}\Big[ \min_{\substack{\tilde{P}_{Z|SUVY}:\\\tilde{P}_{Z|SV}=P_{Z|SV}}}  D(\tilde{P}_{Z|SUVY}\| Q_{Z|Y}) \Big] \bigg]\nonumber\\
&\stackrel{(c)}= \min_{\substack{\tilde{P}_{SUXY}:\\\tilde{P}_{SUX}=P_{SUX}\\\tilde{P}_{SUY}=P_{SUY}}} \bigg[ D(\tilde{P}_{SUXY}\|P_{SU|X}Q_{XY})\nonumber\\&\hspace{0.5cm}+  \mathbb{E}_{{P}_{SVY}}\Big[ \min_{\substack{\tilde{P}_{Z|SVY}:\\\tilde{P}_{Z|SV}=P_{Z|SV}}}  D(\tilde{P}_{Z|SVY}\| Q_{Z|Y}) \Big] \bigg],
 \label{eq:l}
\end{align}
where the steps are justified as follows:
\begin{itemize}
\item [$(a)$] follows because, by the convexity of the KL-divergence, the LHS is larger than or equal to the RHS; the reverse direction holds because the minimization on the LHS can only increase if one restricts pmfs to be of the form $\tilde{P}_{VZ|SUXY}=\tilde{P}_{VZ|SUY}$;
\item [$(b)$] holds because $\tilde{P}_{V|SUY}=P_{V|SUY}$; and
\item [$(c)$]  follows because, by the convexity of the KL-divergence, the LHS is larger than or equal to the RHS; the reverse direction holds because the minimization on the LHS can only increase if one restricts pmfs to be of the form $\tilde{P}_{Z|SUVY}=\tilde{P}_{Z|SVY}$.
\end{itemize}
\medskip
Defining now $\bar{U}:=(U,S)$ and $\bar{V}:=(V,S)$, we conclude that 
\begin{IEEEeqnarray}{rCl}
&&\min_{\substack{\tilde{P}_{SUXY}:\\\tilde{P}_{SUX}=P_{SUX}\\\tilde{P}_{SUY}=P_{SUY}}} D(\tilde{P}_{SUXY}\|P_{SU|X}Q_{XY}) \nonumber\\&& =  \min_{\substack{\tilde{P}_{\bar{U}XY}:\\\tilde{P}_{\bar{U}X}=P_{\bar{U}X}\\\tilde{P}_{\bar{U}Y}=P_{\bar{U}Y}}}  D(\tilde{P}_{\bar{U}XY}\|P_{\bar{U}|X}Q_{XY})
\end{IEEEeqnarray}
and 
\begin{IEEEeqnarray}{rCl}
&&\hspace{-0.5cm}	\min_{\substack{\tilde{P}_{SUVXYZ}:\\\tilde{P}_{SUX}=P_{SUX}\\\tilde{P}_{SUVY}=P_{SUVY}\\\tilde{P}_{SVZ}=P_{SVZ}}} D(\tilde{P}_{SUVXYZ}\|P_{SU|X}P_{V|SUY}Q_{XY}Q_{Z|Y})  \nonumber \\
	&  = &
		\min_{\substack{\tilde{P}_{\bar{U}XY}:\\\tilde{P}_{\bar{U}X}=P_{\bar{U}X}\\\tilde{P}_{\bar{U}Y}=P_{\bar{U}Y}}}  D(\tilde{P}_{\bar{U}XY}\|P_{\bar{U}|X}Q_{XY})\nonumber\\&&\hspace{0.2cm}+  \mathbb{E}_{{P}_{Y}}\Big[ \min_{\substack{\tilde{P}_{\bar{V}Z|Y}:\\\tilde{P}_{\bar{V}Y}=P_{\bar{V}Y}\\\tilde{P}_{\bar{V}Z}=P_{\bar{V}Z}}}  D(\tilde{P}_{\bar{V}Z|Y}\|P_{\bar{V}|Y} Q_{Z|Y}) \Big].
 \end{IEEEeqnarray}
Notice further the Markov chains $\bar{U}\to X \to Y$ and  $\bar{V} \to Y \to Z$ and that the choice $(\bar{U},\bar{V})$ satisfies the rate constraints
\begin{IEEEeqnarray}{rCl}
I(\bar{U};X) = I(S,U;X) \leq R 
\end{IEEEeqnarray}
and 
\begin{IEEEeqnarray}{rCl}
I(\bar{V};Y) &=& I(S;Y) + I(V;Y|S) \nonumber\\&\leq& I(S;X) + I({V};Y,U|S) \nonumber\\ &\leq& T.
\end{IEEEeqnarray}
From all these steps, we  conclude that the choice $P_{\bar{U}|X}=P_{US|X}$ and $P_{\bar{V}|Y}=P_{SV|Y}$ satisfies the following three conditions:
\begin{IEEEeqnarray}{rCl}
I(\bar{U};X)  & \leq &  R \\
I(\bar{V};Y)  & \leq  & T\\
\mathcal{E}_{\simple}(P_{\bar{U}|X}, P_{\bar{V}|Y})  & \supseteq & \mathcal{E}_{\nobinning}(P_{SU|X}, P_{V|SUY}).
\end{IEEEeqnarray}
This proves inclusion \eqref{eq:inc2}.

\section{Proof of Inclusion $\mathcal{E}_{\binsimple} (R,T) \supseteq \mathcal{E}_{\binning}(R,T)$}\label{app:inclusion2}
{Fix a pair of conditional pmfs $P_{SU|X}$ and $P_{V|SUY}$ and define $\bar{U}:=(U,S)$ and $\bar{V}:=(V,S)$. 
Notice  first that, since $\tilde{P}_{SY}=P_{SY}$ and $\tilde{P}_{SZ}=P_{SZ}$, the following hold:
\begin{itemize}
\item Condition $H(U|S,Y) \leq H_{\tilde{P}}(U|S,Y)$ is equivalent to {$H(U,S|Y)    \leq H_{\tilde{P}}(U,S|Y)$ and hence also equivalent to
$H(\bar{U}|Y) \le H_{\tilde{P}}(\bar{U}|Y)$};
\item Condition $H(V|S,Z) \leq H_{\tilde{P}}(V|S,Z)$ is equivalent to {$H(V,S|Z)    \leq H_{\tilde{P}}(V,S|Z)$ and hence also equivalent to $H(\bar{V}|Z) \le H_{\tilde{P}}(\bar{V}|Z)$}.
\end{itemize}
Using these equivalences and following  similar steps as in  the proof of Proposition~\ref{prop:Markov} in Appendix~\ref{app:proof_prop}, it can be shown that}
\begin{IEEEeqnarray}{rCl}
&& \min_{\substack{\tilde{P}_{\bar UXY}:\\\tilde{P}_{\bar  UX}=P_{UX}\\\tilde{P}_{\bar UY}=P_{\bar UY}}} D(\tilde{P}_{\bar UXY}\|P_{\bar U|X}Q_{XY}) \nonumber\\&&\hspace{0.5cm}+  \min_{\substack{\tilde{P}_{\bar VZ|Y}:\\\tilde{P}_{\bar VY}=P_{\bar VY}\\\tilde{P}_{\bar VZ}=P_{\bar VZ}}}  \mathbb{E}_{P_{Y}}\Big[  D(\tilde{P}_{\bar VZ|Y}\|P_{\bar{V}|Y}Q_{Z|Y})\Big]  \nonumber \\[1ex]
&& \geq   \min_{\substack{\tilde{P}_{SUVXYZ}:\\\tilde{P}_{SUX}=P_{SUX}\\\tilde{P}_{SUVY}=P_{SUVY}\\\tilde{P}_{SVZ}=P_{SVZ}}}\!\!\!\!  D(\tilde{P}_{SUVXYZ}\|P_{SU|X}P_{V|SUY}Q_{XYZ});\nonumber\\\\[2ex]
&&\min_{\substack{\tilde{P}_{\bar UXY}:\\\tilde{P}_{\bar UX}=P_{\bar UX}\\\tilde{P}_{Y}=P_{Y}\\H(\bar U|Y)\leq H_{\tilde{P}}(\bar U|Y)}}\!\!\!\!  D(\tilde{P}_{\bar UXY}\|P_{\bar U|X}Q_{XY})\nonumber\\&&\hspace{0.5cm}+  \min_{\substack{\tilde{P}_{\bar VZ|Y}:\\\tilde{P}_{\bar VY}=P_{\bar VY}\\\tilde{P}_{\bar VZ}=P_{\bar VZ}}}  \mathbb{E}_{P_{Y}}\Big[  D(\tilde{P}_{\bar VZ|Y}\|P_{ \bar{V}|Y}Q_{Z|Y})\Big]  \nonumber \\[1ex]
&&
   \geq      \min_{\substack{\tilde{P}_{SUU'VXYZ}:\\\tilde{P}_{SUX}=P_{SUX}\\\tilde{P}_{SVY}=P_{SVY}\\\tilde{P}_{SVZ}=P_{SVZ}\\H(U|S,Y)\leq H_{\tilde{P}}(U|S,Y)}}\!\!\!\!\!\!\!\!\!\! D(\tilde{P}_{SUVXYZ}\| P_{SU|X}P_{V|SY}Q_{XYZ})   ;\nonumber\\\\[2ex]
&&\min_{\substack{\tilde{P}_{\bar UXY}:\\\tilde{P}_{\bar UX}=P_{\bar UX}\\\tilde{P}_{Y}=P_{Y}\\H(\bar U|Y)\leq H_{\tilde{P}}(\bar U|Y)}}\!\!\!\!  D(\tilde{P}_{\bar UXY}\|P_{\bar U|X}Q_{XY}) \nonumber\\*&&\hspace{0.5cm}+ \min_{\substack{\tilde{P}_{\bar VZ|Y}:\\\tilde{P}_{\bar VY}=P_{\bar VY}\\\tilde{P}_{Z}=P_{Z}\\H(\bar V|Z)\leq H_{\tilde{P}}(\bar V|Z)}} \!\!\!\!  \mathbb{E}_{P_{Y}}\Big[  D(\tilde{P}_{\bar VZ|Y}\|P_{\bar V|Y}Q_{Z|Y})\Big]  \nonumber \\[1ex]
&&\geq  \hspace{-3mm}\min_{\substack{\tilde{P}_{SUVXYZ}:\\\tilde{P}_{SUX}=P_{SUX}\\\tilde{P}_{SUVY}=P_{SUVY}\\\tilde{P}_{SZ}=P_{SZ}\\H(V|S,Z)\leq H_{\tilde{P}}(V|S,Z)}} \hspace{-3mm}\!\!\!\!\!\! D(\tilde{P}_{SUVXYZ}\|P_{SU|X}P_{V|SUY}Q_{XYZ});\nonumber\\\\[2ex]    
&&  \min_{\substack{\tilde{P}_{\bar UXY}:\\\tilde{P}_{\bar UX}=P_{\bar UX}\\\tilde{P}_{Y}=P_{Y}\\H(\bar U|Y)\leq H_{\tilde{P}}(\bar U|Y)}}\!\!\!\!  D(\tilde{P}_{\bar UXY}\|P_{\bar U|X}Q_{XY})\nonumber\\&&\hspace{1cm}+
  \min_{\substack{\tilde{P}_{\bar VZ|Y}:\\\tilde{P}_{\bar VY}=P_{\bar VY}\\\tilde{P}_{Z}=P_{Z}\\H(\bar V|Z)\leq H_{\tilde{P}}(\bar V|Z)}} \!\!\!\!  \mathbb{E}_{P_{Y}}\Big[  D(\tilde{P}_{\bar VZ|Y}\|P_{\bar V|Y}Q_{Z|Y})\Big]  \nonumber \\[1ex]
   & \geq 	&  \min_{\substack{\tilde{P}_{SUVXYZ}:\\\tilde{P}_{SUX}=P_{SUX}\\\tilde{P}_{SVY}=P_{SVY}\\\tilde{P}_{SZ}=P_{SZ}\\H(U|S,Y)\leq H_{\tilde{P}}(U|S,Y)\\H(V|S,Z)\leq H_{\tilde{P}}(V|S,Z)}}\!\!\!\! D(\tilde{P}_{SUVXYZ}\| P_{SU|X}P_{V|SY}Q_{XYZ}).\nonumber\\
\end{IEEEeqnarray}
Since moreover
\begin{IEEEeqnarray}{rCl}
- I(\bar{U};X|Y) &=&- I(S,U;X|Y) \nonumber\\&=&- I(S,U;X) +  I(S,U;Y) \nonumber\\&\geq& - I(S,U;X) + I(U;Y|S) \\[1ex]
- I(\bar{V};Y|Z) &= &- I(S,V; Y) + I(S,V;Z) \nonumber\\&\geq & - I(S;Y) - I(V;Y|S) + I(V;Z|S) \nonumber \\
& \geq& - I(S;X) - I(V;U,Y|S) + I(V;Z|S),\nonumber\\
\end{IEEEeqnarray}
we can conclude that 
\begin{equation}
\mathcal{E}_{\binsimple}(P_{\bar U|X}, P_{\bar V|Y}) \supseteq \mathcal{E}_{\binning}(P_{SU|X}, P_{V|SUY}).
\end{equation}
This establishes the desired proof.

\section{Proof of Converse to Corollary \ref{SIthm}}\label{SIproof}

Fix a sequence of encoding and decoding functions $\{\phi^{(n)},\phi_y^{(n)},g_y^{(n)},g_z^{(n)}\}$ so that the inequalities of Definition \ref{cascadedef} hold for sufficiently large blocklengths $n$. Fix also such a sufficiently large $n$ and  define for each $t\in\{1,\ldots, n\}$:
\begin{IEEEeqnarray}{rCl}
	U_t &:{=} & (M,X^{t-1},\YC^{t-1},\YCrest)  \\
	V_t & :{=} & (B,\YH^{t-1},\ZC^{t-1},\ZCrest,\YC^{t-1},\YCrest).
\end{IEEEeqnarray}
Define further $U := (U_T, T)$; $V := (V_T, T)$;  $X :=X_T$; $Y := Y_T$; $W := W_T$; and $Z := Z_T$; for $T\sim\mathcal{U}\{1,\ldots, n\}$ independent of the tuples $(U^n, V^n,  X^n, Y^n,Z^n)$.  Notice the Markov chains $U \to X \to Y$ and $V\to Y \to Z$. Let $\delta(\epsilon):{=}H_{\text{b}}(\epsilon)/(n\cdot (1-\epsilon))$ where $H_{\text{b}}(\epsilon)$ denotes the entropy of the binary random variable with parameter $\epsilon$.  

First, consider the rate $R$:
\begin{align}
R &= \frac{1}{n}H(M)\nonumber\\
&\geq \frac{1}{n}I(M;X^n|\YC^n)\nonumber\\
&=\frac{1}{n}\sum_{t=1}^n I(M;X_t|X^{t-1},\YC^n)\nonumber\\
&\stackrel{(a)}{=}\frac{1}{n}\sum_{t=1}^n I(M,X^{t-1},\YC^{t-1},\YCrest;X_t|\YCt)\nonumber\\
&=\frac{1}{n}\sum_{t=1}^n I(U_t;X_t|\YCt)\nonumber\\
&=I(U;X|\YC),
\end{align}
{where $(a)$ follows from the memoryless property of the sources.}
Similarly, 
\begin{align}
&T = \frac{1}{n}H(B)\nonumber\\
&\geq \frac{1}{n}I(B;Y^n|\YC^n,\ZC^n)\nonumber\\
&= \frac{1}{n}\sum_{t=1}^n I(B;Y_t|Y^{t-1},\YC^n,\ZC^n)\nonumber\\
&=\frac{1}{n}\sum_{t=1}^n I(B,Y^{t-1},\ZC^{t-1},\ZCrest,\YC^{t-1},\YCrest;Y_t|\YCt,\ZCt)\nonumber\\
&=\frac{1}{n}\sum_{t=1}^n I(V_t;Y_t|\YCt,\ZCt)\nonumber\\
&=I(V;Y|\YC,\ZC).
\end{align}
The type-II error probability at the relay can be  bounded as
\begin{align}
&-\frac{1}{n}\log \beta_{y,n}\nonumber\\&\leq \frac{1}{(1-\epsilon)n}D(P_{M\YH^n\YC^n|\mathcal{H}=0}\|P_{M\YH^n\YC^n|\mathcal{H}=1})+\delta(\epsilon)\nonumber\\
&\stackrel{(b)}{=}\frac{1}{(1-\epsilon)n}D(P_{M\YH^n\YC^n}\|P_{M|\YC^n}P_{\YH^n|\YC^n}P_{\YC^n})+\delta(\epsilon)\nonumber\\
&=\frac{1}{(1-\epsilon)n}I(M;\YH^n|\YC^n)+\delta(\epsilon)\nonumber\\
&=\frac{1}{(1-\epsilon)n}\sum_{t=1}^n I(M;\YHt|\YH^{t-1},\YC^n)+\delta(\epsilon)\nonumber\\
&=\frac{1}{(1-\epsilon)n}\sum_{t=1}^n I(M,\YH^{t-1},\YC^{t-1},\YCrest;\YHt|\YCt)+\delta(\epsilon)\nonumber\\
&\stackrel{(c)}{\leq}\frac{1}{(1-\epsilon)n}\sum_{t=1}^n I(M,X^{t-1},\YC^{t-1},\YCrest;\YHt|\YCt)+\delta(\epsilon)\nonumber\\
&=\frac{1}{(1-\epsilon)n}\sum_{t=1}^n I(U_t;\YHt|\YCt)+\delta(\epsilon)\nonumber\\
&=\frac{1}{1-\epsilon}I(U;{Y}|\YC)+\delta(\epsilon),
\end{align}
{where $(b)$ holds by the assumption on the distributions $P_{X\YC\YH\ZC\ZH}$ and $Q_{X\YC\YH\ZC\ZH}$ in \eqref{SI2b}--\eqref{SI2} and the fact that $M$ is a function of $X^n$; and $(c)$ holds by the Markov chain $\YH^{t-1}\to (M,X^{t-1},\YC^n)\to \YHt$.} Finally, consider the type-II error probability at the receiver:
\begin{IEEEeqnarray}{rCl}
\IEEEeqnarraymulticol{3}{l}{-\frac{1}{n}\log \beta_{z,n}} \nonumber\\* &\leq& \frac{1}{(1-\epsilon)n}D(P_{B\ZH^n\ZC^n\YC^n|\mathcal{H}=0}\|P_{B\ZH^n\ZC^n\YC^n|\mathcal{H}=1})+\delta(\epsilon)\nonumber\\ 
&\stackrel{(d)}{=}  &
\frac{1}{(1-\epsilon)n}\mathbb{E}_{\ZC^n\YC^n} \big[D(P_{B\ZH^{n}|\ZC^n\YC^n,\mathcal{H}=0}\|P_{B\ZH^{n}|\ZC^n\YC^n,\mathcal{H}=1})\big]\nonumber\\ &&{}+\delta(\epsilon)\nonumber\\[1ex]
&\stackrel{(e)}{=} & 
\frac{1}{(1-\epsilon)n}\Big(\mathbb{E}_{\ZC^n\YC^n} \big[D(P_{B|\ZC^n\YC^n,\mathcal{H}=0}\|P_{B|\ZC^n\YC^n,\mathcal{H}=1})\big] \nonumber \\
  && \quad{} + \mathbb{E}_{B\ZC^n\YC^n} \big[D(P_{\ZH^{n}|B\ZC^n\YC^n,\mathcal{H}=0}\|P_{\ZH^{n}|B\ZC^n\YC^n,\mathcal{H}=1})\big]\Big)\nonumber\\&&{}+\delta(\epsilon)\nonumber\\[1ex]
 &\stackrel{(f)}{\leq}&\frac{1}{(1-\epsilon)n}\cdot \nonumber\\
 & & \Big(\mathbb{E}_{\ZC^n\YC^n} \big[D(P_{M\YH^n|\ZC^n\YC^n,\mathcal{H}=0}\|P_{M\YH^n|\ZC^n\YC^n,\mathcal{H}=1})\big] \nonumber \\
   && {}+\mathbb{E}_{B\ZC^n\YC^n} \big[D(P_{\ZH^{n}|B\ZC^n\YC^n,\mathcal{H}=0}\|P_{\ZH^{n}|\ZC^n\YC^n, \mathcal{H}=1})\big]\Big)\nonumber\\&&{}+\delta(\epsilon)\nonumber\\[1ex]
& \stackrel{(g)}{=}&\frac{1}{(1-\epsilon)n} \cdot \nonumber\\*
 && \Big(\mathbb{E}_{\YH^n\YC^n\ZC^n} \big[D(P_{M|\YC^n\YH^n\ZC^n,\mathcal{H}=0}\|P_{M|\YC^n\YH^n\ZC^n,\mathcal{H}=1})\big]\nonumber\\&&{}+I(B;\ZH^n|\ZC^n,\YC^n) \Big)+\delta(\epsilon)\nonumber\\[1ex]
 &\stackrel{(h)}{=}&\frac{1}{(1-\epsilon)n}\mathbb{E}_{\YH^n\YC^n} \big[D(P_{M|\YC^n\YH^n,\mathcal{H}=0}\|P_{M|\YC^n\YH^n,\mathcal{H}=1})\big]\nonumber\\&&{}+\frac{1}{(1-\epsilon)n}I(B;\ZH^n|\ZC^n,\YC^n)+\delta(\epsilon)\nonumber\\[1ex]
&\stackrel{(i)}{=}& \frac{1}{(1-\epsilon)n} \Big(I(M; \YH^n|\YC^n) +I(B;\ZH^n|\ZC^n,\YC^n)\Big)+\delta(\epsilon)\nonumber\\[1ex]
&\stackrel{(j)}{=}&\frac{1}{(1-\epsilon)n}  \sum_{t=1}^n\Big[ I\big(M, \YH^{t-1},  \YC^{t-1}, \YCrest;\YHt|\YCt\big) \nonumber \\
 &&{} + I\big(B, \ZH^{t-1}, \ZC^{t-1}, \ZCrest, \YC^{t-1}, \YCrest ;\ZHt \; \big| \nonumber\\* && \qquad\qquad\qquad\qquad\qquad\qquad\qquad \ZCt,\YCt)\Big]+\delta(\epsilon)\nonumber\\[1ex]
&\stackrel{(k)}{\leq}& \frac{1}{(1-\epsilon)n} \sum_{t=1}^n I(U_t;\YHt|\YCt)\nonumber\\&&{}+ \frac{1}{(1-\epsilon)n} \sum_{t=1}^{n} I(V_t;\ZHt| \ZCt,\YCt)+\delta(\epsilon)\nonumber\\
&=& \frac{1}{1-\epsilon}I(U;Y|\YC)+ \frac{1}{1-\epsilon}I(V;Z|\ZC,\YC)+\delta(\epsilon),
\end{IEEEeqnarray}
{where $(d)$ holds because the pair $(\YC^n,\ZC^n)$ has the same distribution under both hypotheses; $(e)$ holds by the chain rule for KL-divergence; $(f)$ holds by the data-processing inequality and the fact that $B$ is a function of $(M,\YH^n,\YC^n)$, and because under $\mathcal{H}=1$ and   given $(\YC^n,\ZC^n)$, the message $B$ is independent of the observation  $\ZH^n$; $(g)$ holds because the two triples  $(\YH^n,\YC^n,\ZC^n)$ and $(\YC^n, \ZH^n,\ZC^n)$ have the same distribution under both hypotheses; $(h)$ holds because under both hypotheses $M$ is independent of $\ZH^n$ given the pair $(\YH^n,\YC^n)$; $(i)$ holds because the triple $(M,\YH^n,\YC^n)$ has same distribution under both hypotheses; $(j)$ holds by the memoryless property of the sources;  and $(k)$ holds by the definitions of $U_t$ and $V_t$ and the  Markov chain $\ZH^{t-1}\to (B,\YH^{t-1},\ZC^n,\YC^n)\to \ZHt$.}

\section{Proof of the Converse to corollary~\ref{prop1b}}\label{app:convCor}
 Fix  sequences of encoding and decoding functions $\{\phi^{(n)},\phi_y^{(n)},g_y^{(n)},g_z^{(n)}\}$, and notice that there exists a function $\delta(\epsilon)$ which tends to zero when $\epsilon\to 0$ such that, for any $\epsilon>0$ and sufficiently large $n$: 
	\begin{IEEEeqnarray}{rCl}
		\IEEEeqnarraymulticol{3}{l}{-\frac{1}{n}\log \beta_{y,n}} \nonumber\\* \;\;\;\; & \leq&  \frac{1}{(1-\epsilon)n}D(P_{MY^n|\mathcal{H}=0}\|P_{MY^n|\mathcal{H}=1})+\delta(\epsilon)\nonumber\\&= &\delta(\epsilon) \nonumber\\
		\IEEEeqnarraymulticol{3}{l}{-\frac{1}{n}\log \beta_{z,n}} \nonumber\\* &  \leq &\frac{1}{(1-\epsilon)n} D(P_{B\ZC^n\ZH^n|\mathcal{H}=0}\|P_{B\ZC^n\ZH^n|\mathcal{H}=1})\nonumber\\&& {}+ \delta(\epsilon) \nonumber\\
		&  \stackrel{(a)}{=}& \frac{1}{(1-\epsilon)n} D(P_{B\ZH^n|\ZC^n,\mathcal{H}=0}\|P_{B\ZH^n|\ZC^n,\mathcal{H}=1})+ \delta(\epsilon) \nonumber\\
		&\stackrel{(b)}{=} &\frac{1}{(1-\epsilon)n}  I(B;\ZH^n|\ZC^n) + \delta(\epsilon) \nonumber \\
		& = & \frac{1}{(1-\epsilon)n}\sum_{t=1}^{n} I(B, \ZH^{t-1}; \ZHt | \ZC^n)+ \delta(\epsilon) \nonumber\\
		& = & \frac{1}{(1-\epsilon)n}\sum_{t=1}^{n} I(B, \ZH^{t-1},  \ZC^{t-1},\ZCrest; \ZHt | \ZCt)\nonumber\\&&{}+ \delta(\epsilon) \nonumber\\
		&\stackrel{(c)}{\leq}& \frac{1}{(1-\epsilon)n}\sum_{t=1}^{n} I(B, Y^{t-1}, \ZCrest; \ZHt| \ZCt)+ \delta(\epsilon),\nonumber
	\end{IEEEeqnarray}
 where  {$(a)$ holds because $\ZC^n$ has the same distribution under both hypotheses; $(b)$ holds because, {conditional on $\ZC^n$, the two random variables $B$ and $\ZH^n$ have the same marginals under both hypothesis, while being dependent under $\mathcal{H}=0$ and independent under $\mathcal{H}=1$}; $(c)$ holds by the Markov chain $ \ZHt \to (B,Y^{t-1},\ZCt^n) \to (\ZC^{t-1},\ZH^{t-1})$.} Moreover,
\begin{align}
T= \frac{1}{n}H(B)&\geq  \frac{1}{n} I(B;Y^n|\ZC^n) \nonumber\\&= \frac{1}{n}\sum_{t=1}^n I(B,Y^{t-1}, \ZC^{t-1}, \ZCrest;Y_t|\ZCt)\nonumber\\&{ \stackrel{(d)}{=} } \frac{1}{n}\sum_{t=1}^n I(B,Y^{t-1},  \ZCrest;Y_t|\ZCt),
\end{align}
where {$(d)$ holds by the Markov chain $ Y_t \to ( B, \ZCt^n,Y^{t-1}) \to \ZC^{t-1}$.} The proof is finalized by introducing auxiliary random variables $V_t:= (B,Y^{t-1}, \ZCrest)$, $t\in\{1,\ldots,n\}$, relabeling the random variables, and taking $\epsilon \to 0$.

\section{Proof of Proposition \ref{thm4}}\label{convthm4}
We fix a sufficiently large $n$ and a sequence of encoding and decoding functions such that the properties of Definition \ref{cascadedef} hold. Also,
define $S_t:{=}(M,X^{t-1},Z^{t-1})$. Notice the Markov chain $S_t \to X_t \to (Y_t,Z_t)$. First, consider the rate $R$:
\begin{align}
nR &= H(M)\nonumber\\
&\geq I(M;X^n,Z^n)\nonumber\\
&=\sum_{t=1}^n I(M;X_t,Z_t|X^{t-1},Z^{t-1})\nonumber\\
&\stackrel{(a)}{=}\sum_{t=1}^n I(M,X^{t-1},Z^{t-1};X_t,Z_t)\nonumber\\
&\geq \sum_{t=1}^n I(M,X^{t-1},Z^{t-1};X_t)\nonumber\\
&=\sum_{t=1}^n I(S_t;X_t),\nonumber
\end{align}
{where $(a)$ holds by the memoryless property of the sources.}
Now, consider the error exponent at the relay. We have:
\begin{align}
-\frac{1}{n}\log \beta_{y,n} &\leq \frac{1}{(1-\epsilon)n} D(P_{MY^n|\mathcal{H}=0}\|P_{MY^n|\mathcal{H}=1})+ \delta(\epsilon)\nonumber\\
&\stackrel{(b)}{=} \frac{1}{(1-\epsilon)n}  I(M;Y^n)+\delta(\epsilon)\nonumber\\
&= \frac{1}{(1-\epsilon)n}  \sum_{t=1}^n I(M;Y_t|Y^{t-1})+\delta(\epsilon)\nonumber\\
&\stackrel{(c)}{=}  \frac{1}{(1-\epsilon)n}   \sum_{t=1}^n I(M,Y^{t-1};Y_t)+\delta(\epsilon)\nonumber\\
&\leq  \frac{1}{(1-\epsilon)n}  \sum_{t=1}^n I(M,X^{t-1},Y^{t-1},Z^{t-1};Y_t)\nonumber\\&\hspace{0.5cm}+\delta(\epsilon)\nonumber\\
&\stackrel{(d)}{=} \frac{1}{(1-\epsilon)n}    \sum_{t=1}^n I(M,X^{t-1},Z^{t-1};Y_t)+\delta(\epsilon)\nonumber\\
&= \frac{1}{(1-\epsilon)n}   \sum_{t=1}^n I(S_t;Y_t)+\delta(\epsilon)\nonumber
\end{align}
where $(b)$ holds because under hypothesis $\mathcal{H}=1$, the message $M$ and the observation $Y^n$ are independent; $(c)$ holds by the memoryless property of the sources; and $(d)$ by the Markov chain $(Y^{t-1},Z^{t-1})\to (M,X^{t-1})\to Y_t$.
Next, consider the error exponent at the receiver:
\begin{align}
-\frac{1}{n}\log \beta_{z,n} &\leq \frac{1}{(1-\epsilon)n}  D(P_{BZ^n|\mathcal{H}=0}\|P_{BZ^n|\mathcal{H}=1})+\delta(\epsilon)\nonumber\\
&\stackrel{(e)}{\leq} \frac{1}{(1-\epsilon)n}  D(P_{MY^nZ^n|\mathcal{H}=0}\|P_{MY^nZ^n|\mathcal{H}=1})\nonumber\\&\hspace{0.5cm}+\delta(\epsilon)\nonumber\\
&\stackrel{(f)}{=}\frac{1}{(1-\epsilon)n} I(M;Y^n,Z^n)+\delta(\epsilon)\nonumber\\
&\stackrel{(g)}{=}\frac{1}{(1-\epsilon)n} I(M;Z^n)+\delta(\epsilon)\nonumber\\
&=\frac{1}{(1-\epsilon)n} \sum_{t=1}^nI(M;Z_t|Z^{t-1})+\delta(\epsilon)\nonumber\\
&=\frac{1}{(1-\epsilon)n} \sum_{t=1}^nI(M,Z^{t-1};Z_t)+\delta(\epsilon)\nonumber\\
&\leq \frac{1}{(1-\epsilon)n} \sum_{t=1}^nI(M,X^{t-1},Z^{t-1};Z_t)+\delta(\epsilon)\nonumber\\
&=\frac{1}{(1-\epsilon)n} \sum_{t=1}^nI(S_t;Z_t)+\delta(\epsilon)\nonumber
\end{align}
where $(e)$ holds by the data processing inequality and because $B$ is a function of $M$ and $Y^n$;  $(f)$ holds because $M$ and $(Y^n,Z^n)$ are independent under hypothesis $\mathcal{H}=1$ with same marginals as under $\mathcal{H}=0$; and $(g)$ holds by the Markov chain $M\to X^n\to Z^n\to Y^n$. The proof of the converse is finally concluded by defining a time-sharing random variable $Q\sim\mathcal{U}\{1,...,n\}$ and $S :{=} (S_Q,Q)$, $X:{=} X_Q$,  $Y:{=} Y_Q$ and $Z:{=} Z_Q$ and letting $\epsilon \to 0$ and $n\to \infty$.

\begin{IEEEbiographynophoto}{Sadaf Salehkalaibar} (S'10--M'14) received the B.Sc., M.Sc. and Ph.D.  degrees in Electrical Engineering from Sharif University of Technology, Tehran, Iran in 2008, 2010 and 2014, respectively. She was a postdoctoral fellow at Telecom ParisTech, Paris, France in 2015 and 2017. She is currently an assistant professor at Electrical and Computer Engineering Department of University of Tehran, Tehran, Iran. Her special fields of interest include network information theory, hypothesis testing and fundamental limits of secure communication with emphasis on information-theoretic security.\end{IEEEbiographynophoto}

\begin{IEEEbiographynophoto}{Mich\`ele Wigger}   (S'05, M'09, SM'14) received the M.Sc. degree in electrical
	engineering, with distinction, and the Ph.D. degree in electrical engineering
	both from ETH Zurich in 2003 and 2008, respectively. In 2009, she was
	first a post-doctoral fellow at the University of California, San Diego, USA,
	and then joined Telecom Paris Tech, Paris, France, where she is currently a Full Professor. Dr. Wigger has held visiting professor appointments at
	the Technion-Israel Institute of Technology and ETH Zurich. Dr. Wigger has
	previously served as an Associate Editor of the IEEE Communication Letters,
	and is now Associate Editor for Shannon Theory of the IEEE Transactions on
	Information Theory. She is currently also serving on the Board of Governors
	of the IEEE Information Theory Society. Dr. Wigger's research interests are in
	multi-terminal information theory, in particular in distributed source coding
	and in capacities of networks with states, feedback, user cooperation, or
	caching.\end{IEEEbiographynophoto}

\begin{IEEEbiographynophoto}{Ligong Wang}  (S'08--M'12) received the B.E. degree in electronic engineering
	from Tsinghua University, Beijing, China, in 2004, and the M.Sc. and Dr.Sc.
	degrees in electrical engineering from ETH Zurich, Switzerland, in 2006 and
	2011, respectively. In the years 2011--2014 he was a Postdoctoral Associate
	at the Department of Electrical Engineering and Computer Science at the
	Massachusetts Institute of Technology, Cambridge, MA, USA. He is now a
	researcher (chargé de recherche) with CNRS, France, and is affiliated with
	ETIS laboratory in Cergy-Pontoise. His research interests include classical and
	quantum information theory, physical-layer security, and digital, in particular
	optical communication.  \end{IEEEbiographynophoto}

\end{document}